\documentclass[journal]{IEEEtran}
\usepackage{amsmath,amsfonts,amsthm}
\usepackage[ruled,commentsnumbered,linesnumbered]{algorithm2e}
\usepackage{array}
\usepackage{textcomp}
\usepackage[dvipsnames]{xcolor}
\usepackage{stfloats}
\ifCLASSOPTIONcompsoc
\usepackage[caption=false,font=normalsize,labelfont=sf,textfont=sf]{subfig}
\else
\usepackage[caption=false,font=footnotesize]{subfig}
\fi
\usepackage{url}
\usepackage{verbatim}
\usepackage{graphicx}
\usepackage{cite}
\usepackage[colorlinks,linkcolor=BrickRed,citecolor=Aquamarine,urlcolor=cyan]{hyperref} % blue
\usepackage{makecell}
\usepackage{multirow}
\usepackage{color}
\usepackage{booktabs}
\usepackage{amsmath}

\allowdisplaybreaks[4]
\usepackage{amssymb}
\usepackage{threeparttable}

\theoremstyle{remark}
\newtheorem{remark}{Remark}
\newtheorem{proposition}{Proposition}
\newtheorem{lemma}{Lemma}
\newtheorem{definition}{Definition}
\newtheorem{case}{Case}
\newtheorem{assumption}{Assumption}
\newtheorem{theorem}{Theorem}
\newtheorem{corollary}{Corollary}
\newtheorem{example}{Example}

\usepackage{orcidlink} 

\begin{document}

\title{Finite-Precision Arithmetic Transceiver for Massive MIMO Systems}
\author{Yiming Fang$^{\orcidlink{0009-0003-6864-6422}}$,~\IEEEmembership{Graduate Student Member,~IEEE}, 
    Li Chen$^{\orcidlink{0000-0002-1754-0607}}$,~\IEEEmembership{Senior Member,~IEEE}, \\
    Yunfei Chen$^{\orcidlink{0000-0001-8083-1805}}$,~\IEEEmembership{Senior Member,~IEEE}, 
    and Huarui Yin$^{\orcidlink{0000-0003-0506-5930}}$,~\IEEEmembership{Member,~IEEE}
        % <-this % stops a space
    \thanks{This work was supported in part by the Natural Science Foundation of China (Grant No. 62471449) and Anhui Provincial Natural Science Foundation (No.  2308085J24). Part of this paper is presented at the IEEE Global Communications Conference (GLOBECOM), Cape Town, South Africa, 2024.}
    \thanks{Yiming Fang, Li Chen, and Huarui Yin are with the CAS Key Laboratory of Wireless-Optical Communications, University of Science and Technology of China, Hefei 230027, China (e-mail: fym1219@mail.ustc.edu.cn; \{chenli87, yhr\}@ustc.edu.cn).}
    \thanks{Yunfei Chen is with the Department of Engineering, University of Durham, Durham, UK, DH1 3LE (e-mail: Yunfei.Chen@durham.ac.uk).}
}

\markboth{IEEE Journal on Selected Areas in Communications}
{Fang \MakeLowercase{\textit{et al.}}: Finite-Precision Arithmetic Transceiver for Massive MIMO Systems}


\maketitle
\begin{abstract}
Efficient implementation of massive multiple-input-multiple-output (MIMO) transceivers is essential for the next-generation wireless networks. To reduce the high computational complexity of the massive MIMO transceiver, in this paper, we propose a new massive MIMO architecture using finite-precision arithmetic. First, we conduct the rounding error analysis and derive the lower bound of the achievable rate for single-input-multiple-output (SIMO) using maximal ratio combining (MRC) and multiple-input-single-output (MISO) systems using maximal ratio transmission (MRT) with finite-precision arithmetic. Then, considering the multi-user scenario, the rounding error analysis of zero-forcing (ZF) detection and precoding is derived by using the normal equations (NE) method. The corresponding lower bounds of the achievable sum rate are also derived and asymptotic analyses are presented. Built upon insights from these analyses and lower bounds, we propose a mixed-precision architecture for massive MIMO systems to offset performance gaps due to finite-precision arithmetic. The corresponding analysis of rounding errors and computational costs is obtained. Simulation results validate the derived bounds and underscore the superiority of the proposed mixed-precision architecture to the conventional structure.
\end{abstract}


\section{Introduction}
\label{sec:intro}
\IEEEPARstart{M}{assive} multiple-input-multiple-output (MIMO) can greatly enhance the energy and spectral efficiencies for the next-generation wireless networks \cite{6457363}. Nevertheless, in practical implementation, the complexity of transceiver design becomes unbearable due to the substantial number of antennas \cite{SUM}. Thus, low-complexity transceiver design is desirable and has recently drawn significant attention.

To alleviate the hardware complexity of high-resolution analog-to-digital converter (ADC), low-resolution ADC is often employed. In the exploration of the impact of coarse quantization on massive MIMO systems with low-resolution ADCs, various theoretical frameworks, such as the Bussgang decomposition \cite{9307295} and generalized mutual information (GMI) \cite{7437384}, have been utilized. Moreover, advanced signal detectors based on message passing \cite{8917914} and Bayesian inference \cite{7355388} have been proposed for mitigating the impact of nonlinear distortion resulting from quantization. To reduce the number of radio frequency (RF) chains, the utilization of hybrid beamforming architecture has also been widely investigated. It has been found to achieve good spectral efficiency while maintaining low hardware complexity \cite{7389996}. Furthermore, the authors in \cite{9374093} proposed a dynamic hybrid (DH) architecture with a switch network, demonstrating higher energy efficiency. Moreover, low-resolution adjustable phase shifters were utilized for the DH architecture in \cite{9110865}. Additionally, the authors in \cite{9934006} adopted fixed phase shifters (FPS) to reduce power consumption associated with the phase selection process. 

In addition to the hardware complexity, the computational complexity of the transceiver has also garnered significant attention. This is because the increasing number of BS antennas leads to the high dimensions of matrix computations in the transceiver realization. To address this problem, there has been notable interest in the application of graphics processing units (GPUs), which can accelerate the computation in massive MIMO systems. The authors in \cite{9083670} efficiently solved the classical proportional-fair (PF) scheduling problem using GPU. Additionally, the authors in \cite{9716803} and \cite{10329960} proposed GPU-based solutions for hybrid beamforming design and multi-cell MIMO scheduling, respectively. Considering the analog computing architecture, the authors in \cite{wang2023parallel} utilized memristor-based analog circuits to validate massive MIMO orthogonal frequency division multiplexing (OFDM) transceivers. Furthermore, a fully parallel memristor-based circuit was proposed for minimum mean squared error (MMSE) detection in \cite{10273392}. In \cite{10092945}, photonic computing was utilized to expedite data processing in massive MIMO systems.

Moreover, the exploitation of channel properties is also promising to reduce the computation complexity of the transceiver realization. The authors in \cite{FMI} proposed a low-complexity zero-forcing (ZF) precoding using the channel state information from the previous time slot. Considering nonlinear precoding, low-complexity Tomlinson-Harashima precoding (THP) algorithms were presented in \cite{10368038} based on the temporal correlation of the channel. Additionally, the authors in \cite{9647981} proposed low-complexity precoding methods and beam selection schemes by leveraging the spatial sparsity of the channel.

The aforementioned works on complexity reduction commonly assume that the matrix computations are carried out in full-precision arithmetic\footnotemark{}\footnotetext{In this paper, we consider full-precision arithmetic to be equivalent to a 64-bit or higher floating-point number format and finite-precision arithmetic to be equivalent to a 32-bit or lower floating-point number format.}. However, finite-precision or low-precision arithmetic could offer further advantages in reducing the computational complexity compared with full-precision arithmetic \cite{higham2022mixed}. For example, half-precision arithmetic outperforms double-precision not only due to its approximately fourfold faster processing but also because half-precision data necessitates only a quarter of the storage and incurs a quarter of the memory transfer costs of the double-precision data.

%Therefore, in this paper, we discuss the computational complexity of massive MIMO transceiver realization through a novel finite-precision arithmetic perspective. 
Only a few works have investigated reducing the computational complexity of communication systems using finite-precision arithmetic. The authors in \cite{10146452} utilized low-precision Cholesky decomposition to implement efficient symbol detection, and the corresponding rounding error was analyzed in \cite{10293148}. However, to the best of our knowledge, the realization of low computational complexity transceivers taking advantage of the finite-precision arithmetic has never been studied. The challenges are twofold. First, the impact of finite-precision arithmetic on communication performance remains unknown. Second, how to alleviate the performance gap due to finite-precision arithmetic and realize a massive MIMO transceiver that approaches the performance of full-precision arithmetic is also an open issue.

To fill in this gap, in this paper, we propose a massive MIMO transceiver from a finite-precision arithmetic perspective. First, we derive the rounding error bound and lower bound of the achievable rate for single-input-multiple-output (SIMO) systems using maximal ratio combining (MRC) and multiple-input-single-output (MISO) systems using maximal ratio transmission (MRT) with finite-precision arithmetic. Then, we extend these results to general multi-user scenarios, i.e., multi-user SIMO (MU-SIMO) and multi-user MISO (MU-MISO). Finally, to compensate for the performance gap resulting from finite-precision arithmetic, a mixed-precision arithmetic architecture is proposed, and the corresponding analyses of rounding errors and computational costs are presented. Our main contributions are summarized as follows. % Additionally, we propose a mixed-precision arithmetic transceiver architecture that amalgamates the advantages of low-precision and high-precision arithmetic.
\begin{itemize}
    \item \textbf{Impact of finite-precision arithmetic for SIMO and MISO systems.} To acquire fundamental insights into finite-precision arithmetic, the special case of a single user is first studied. We derive the rounding error bound and the lower bound of the achievable rate for SIMO and MISO systems with finite-precision arithmetic, which demonstrates the impact of the number of base station (BS) antennas, signal-to-noise ratio (SNR), and the precision of arithmetic. More importantly, we reveal some intriguing observations. Specifically, increasing the BS antennas is not always beneficial for SIMO; a floor effect of the rounding errors exists for both SIMO and MISO systems, and duality between SIMO and MISO does not hold with finite-precision arithmetic.
    \item \textbf{Extension of finite-precision arithmetic to the multi-user scenario.} For the general multi-user scenario, to eliminate inter-user interference, we utilize ZF detection for MU-SIMO systems and ZF precoding for MU-MISO systems, respectively. Note that ZF is engaged in matrix inversion, rendering the analysis of its rounding error challenging. Hence, we transform ZF into the least squares (LS) problem and propose the normal equations (NE) method-based precoding and detection. Then, the rounding error bounds and the lower bounds of the achievable sum rate are derived. These bounds are explicit functions of BS antennas, SNR, channel conditions, arithmetic precision, and the number of users. Moreover, asymptotic analyses are presented by our derived bounds.
    \item \textbf{Mixed-precision arithmetic transceiver architecture for massive MIMO systems.} To address the performance degradation induced by finite-precision arithmetic, especially low-precision arithmetic, we propose a mixed-precision arithmetic architecture for massive MIMO systems. Specifically, this approach involves partitioning summations in matrix computations into blocks, where intra-block partial sums are computed in low-precision arithmetic and then inter-block sums in high-precision arithmetic. Furthermore, a comprehensive analysis of rounding errors and computational costs is obtained to show the superiority of the proposed architecture.
\end{itemize}

\textit{Organization:} Sec. \ref{sec:sys} describes a multi-user massive MIMO system model and introduces the basic preliminaries of finite-precision arithmetic. In Sec. \ref{sec:su}, we derive the lower bound of the achievable rate for SIMO and MISO systems with finite-precision arithmetic. In Sec. \ref{sec:mu}, the analysis of finite-precision arithmetic is extended to MU-SIMO and MU-MISO systems. Mixed-precision arithmetic transceiver architecture and corresponding performance analysis are presented in Sec. \ref{sec:com}. Numerical results are presented in Sec. \ref{sec:sim}, and the conclusions are provided in Sec. \ref{sec:con}.

\textit{Notation:} Bold uppercase letters denote matrices and bold lowercase letters denote vectors. For a matrix $\bf A$, ${\bf A}^T$, ${\bf A}^H$ and ${\bf A}^{-1}$ denote the transpose, the Hermitian transpose and inverse of ${\bf A}$, respectively. ${a}_{i,j}$ denotes $(i,j)$th entry of ${\bf A}$. $\mathrm{tr}(\bf A)$ denotes the trace of matrix ${\bf A}$. $\mathbb{E}\{\bf A\}$ denotes the expectation of $\bf A$. $ \left| {\bf A}\right|$ represents the matrix of absolute values, $(\left| {a}_{i,j}\right|)$. $\left\| {\bf A} \right\|_2$ denotes its spectral norm. $\kappa_2({\bf A})=\left\| {\bf A} \right\|_2\left\| {\bf A}^{-1} \right\|_2$ represents the condition number of ${\bf A}$. For a vector $\bf a$, $\left\| \bf a \right\|_2$ denotes its Euclidean norm. The notations $\mathbb{R}$ and $\mathbb{C}$ represent the sets of complex numbers and real numbers, respectively. $\Re\{x\}$ and $\Im\{x\}$ denote the real part and imaginary part of $x$. $\lceil x \rceil$ and $\lfloor x \rfloor$ represent the smallest integer more than $x$ and the largest integer no more than $x$, respectively.

\section{System Model}
\label{sec:sys}
\subsection{System Description}
\label{sec:sys_de}
Consider a multi-user massive MIMO system with a BS equipped with $M$ antennas, serving $K$ single-antenna users, where $M\gg K$. Both the BS and users are perfectly synchronized. We assume a flat fading channel where the elements of the channel matrix are modeled as independent complex Gaussian random variables with a zero mean and unit variance.
\subsubsection{Uplink} In the uplink, all the users transmit their signals to the BS in the same time-frequency block. The received signals $\bf z$ at the BS can be written as 
\begin{equation}
\label{eq:up_all}
    {\bf z} = \sqrt{\rho_\mathrm{u}}{\bf H}{\bf x}^{\mathrm{u}} + {\bf n},
\end{equation}
where ${\bf H}=\left[{\bf h}_1,{\bf h}_2,\cdots,{\bf h}_k\right]\in \mathbb{C}^{M\times K}$ is the channel matrix, ${\bf h}_{k}\in \mathbb{C}^{M\times 1}$ is the channel between the BS and the $k$th user, ${\bf x}^{\mathrm{u}}\sim \mathcal{CN}(0,{\bf I}_K)$ is the transmitted signals from all the users, ${\bf n}\sim \mathcal{CN}(0,{\bf I}_M)$ is the additive white Gaussian noise (AWGN), and ${\rho_\mathrm{u}}$ is the average transmitted power of each user, i.e., the uplink SNR when the noise power is 1. 

Then, we utilize a linear detection matrix $\mathbf{A} \in \mathbb{C}^{M\times K}$ to recover the transmitted signals, resulting in
\begin{equation}
\label{eq:up_linear}
    {\bf r} = {\bf A}^{H}{\bf z}.
\end{equation}
Substituting \eqref{eq:up_all} into \eqref{eq:up_linear}, yields
\begin{equation}
\label{eq:up_linear_mu}
    {\bf r} = \sqrt{\rho_\mathrm{u}}{\bf A}^{H}{\bf H}{\bf x}^{\mathrm{u}} + {\bf A}^{H}{\bf n}.
\end{equation}
Let $x_k^\mathrm{u}$ and $r_k$ be the $k$th elements of ${\bf x}^{\mathrm{u}}$ and ${\bf r}$, respectively. Then the received signal for the $k$th user after detecting at the BS can be expressed as 
\begin{equation}
\label{eq:up_linear_su}
    r_k = \sqrt{\rho_\mathrm{u}}{\bf a}_k^{H}{\bf h}_k{x}_k^\mathrm{u} + \sqrt{\rho_\mathrm{u}}\sum_{i=1,i\neq k}^{K}{\bf a}_k^{H}{\bf h}_i{x}_i^\mathrm{u} +{\bf a}_k^{H}{\bf n},
\end{equation}
where ${\bf a}_k$ is the $k$th column of $\bf A$.
\subsubsection{Downlink} In the downlink, the received signals ${\bf y}$ at the users side can be expressed as 
\begin{equation}
\label{eq:dl_linear_mu}
    {\bf y} = \sqrt{\rho_\mathrm{d}}{\bf H}^{H}{\bf s} + {\bf n},
\end{equation}
where ${\bf s}\in \mathbb{C}^{M\times 1}$ is the transmitted vector after precoding at the BS, and $\rho_\mathrm{d}$ is the total transmit power of the BS, i.e., the downlink SNR. Without loss of generality, we denote $\rho = \rho_\mathrm{d} = \rho_\mathrm{u}$. The transmitted vector ${\bf s}$ is given by
\begin{equation}
\label{eq:precoding}
   {\bf s} = \sqrt{\beta}{\bf P}{\bf x}^{\mathrm{d}},
\end{equation}
where ${\bf x}^{\mathrm{d}}\sim \mathcal{CN}(0,{\bf I}_K)$ is the transmitted signals from BS, ${\bf P}\in \mathbb{C}^{M\times K}$ is a linear precoding matrix and $\beta = K/{\mathbb{E}\{\mathrm{tr}({\bf P}{\bf P}^H)\}}$ is normalization factor \cite{6415388}.

Then the received signal at the $k$th user is expressed by
\begin{equation}
\label{eq:dl_linear_su}
    y_k = \sqrt{\rho\beta}{\bf h}_k^{H}{\bf p}_k x_k^\mathrm{d} + \sqrt{\rho\beta}\sum_{i=1,i\neq k}^{K}{\bf h}_k^{H}{\bf p}_i x_i^\mathrm{d} + n_k,
\end{equation}
where ${\bf p}_k$ is the $k$th column of $\bf P$.

 \begin{figure}[t]
     \centering
     \includegraphics[width=0.4\textwidth]{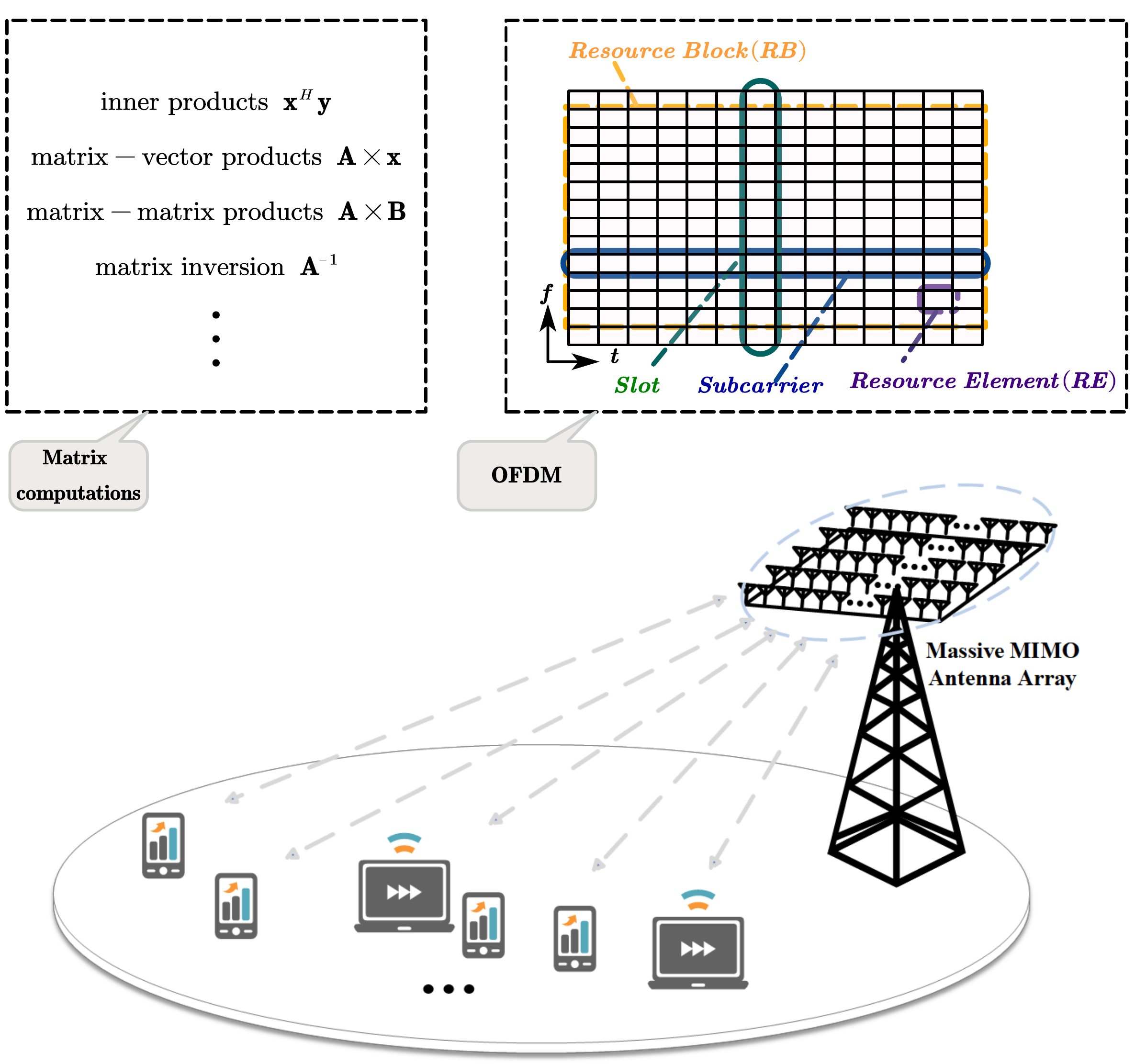}
     \caption{The illustration of the system model, the involved matrix computations, and MIMO-OFDM time-frequency grid.}
     \label{fig:ofdm}
 \end{figure}
 
% In traditional communication systems analysis, computations, such as inner or matrix-matrix products, are usually assumed to occur in full-precision arithmetic\footnote{In this paper, we consider full-precision arithmetic to be equivalent to a 64-bit or higher floating-point number format and finite-precision arithmetic to be equivalent to a 32-bit or lower floating-point number format.}. However, in practical scenarios, finite-precision arithmetic is always involved, and it is crucial to analyze the effects of finite-precision arithmetic.

As shown in Fig. \ref{fig:ofdm}, matrix computations at BS with massive antenna array, such as inner or matrix-matrix products, i.e., \eqref{eq:up_linear} and \eqref{eq:precoding}, are usually assumed to occur in full-precision arithmetic, which leads to huge computational complexity.  We give the following example to further elaborate on this challenge.
 
\begin{example}
We consider a massive MIMO system employing full-precision arithmetic, where the BS has over a thousand antennas, e.g., $M = 1024$, serving more than a hundred single-antenna users, e.g., $K = 100$. For ZF detection or precoding in such a system, computational complexity is $\mathcal{O}\left(MK^2+K^3\right)$ \cite{9647981}, translating to an estimated computational cost of approximately 10 million floating-point operations (flops) per subchannel. Furthermore, considering an OFDM system supporting a maximum of 3300 subcarriers and a slot duration of $30\mathrm{\mu s}$ \cite{8412469}, the required flops rate for completing the detection or precoding is on the order of 1000 tera floating-point operations per second (TFLOPS). Notably, the most potent GPU architecture currently available, NVIDIA V100, boasting 640 tensor cores and 21 billion transistors, can deliver over 130 TFLOPS \cite{nvidia}. In other words, the necessity for multiple GPU units to meet the computational demand results in an untenable increase in both computational cost and power consumption.
\end{example}
Compared with full-precision arithmetic, finite-precision or low-precision arithmetic incurs significantly lower storage and memory transfer costs, and it also exhibits faster computational speeds. For instance, single-precision data requires half as much storage as double-precision data and has half the memory transfer costs. Consequently, a natural consideration arises: \textit{Can we replace full-precision arithmetic with finite-precision or low-precision arithmetic in communication systems?} Next, we provide a fundamental introduction to floating-point arithmetic and rounding error analysis. %

\subsection{Floating-Point Arithmetic and Rounding Error Analysis}
First, we recall that a floating-point number system $\mathbb{F}$ is a subset of real numbers whose elements can be expressed as \cite{higham2022mixed}
\begin{equation}
\label{eq:fp_representation}
    f = \pm m\times \eta^{e-t+1},
\end{equation}
where $\eta = 2$ is the base, the integer $t$ is the precision, the integer $e$ is the exponent within the range $e_{\min}\leq e\leq e_{\max}$, and the integer $m$ is significand satisfying $0\leq m \leq \eta^t - 1$. Table \ref{tab:parafp} provides parameters for four floating-point arithmetic according to the IEEE standard \cite{4610935}.

Then, to delve into rounding error analysis, we present essential definitions and assumptions of basic floating-point arithmetic below.
\begin{definition}[\textit{Floating-point operator}]
\label{def:fp_op}
    $\boldsymbol{fl}\left(\cdot\right)$ is the operator that rounds a real number into the floating-point number system $\mathbb{F}$ whose elements are given by \eqref{eq:fp_representation}.
\end{definition}

\begin{definition}[\textit{Standard arithmetic model}]
\label{def:samodel}
    Denote by $u$ the \textit{unit roundoff}. The floating-point system $\mathbb{F}$ adheres to a standard arithmetic model if, for any $x, y$ is in the range of $\mathbb{F}$. One has{\cite[Sec. 2.2]{higham2002accuracy}}
    \begin{equation}
    \label{eq:model}
        \boldsymbol{fl}\left(x\,\mathrm{op}\,y\right) = \left(x\,\mathrm{op}\,y\right)(1+\delta),
    \end{equation}
    where $\delta \in \mathbb{R}$ is such that $\left|\delta\right|\leq u$, for $\mathrm{op}=+,-,\times,/$.
\end{definition}
Note that the model in \textit{Definition} \ref{def:samodel} is satisfied by the IEEE arithmetic standard\footnote{This paper disregards underflow or overflow impacts, focusing on exploring the effects of low-precision arithmetic independent of range limitations on communication systems.}\cite{higham2022mixed}. Various \textit{unit roundoffs} $u$ in Table \ref{tab:parafp} can be chosen to represent distinct precision arithmetic.

\begin{table}[t]
\centering 
\setlength{\tabcolsep}{3.5pt}
\caption{Parameters for Four Floating-point Arithmetic}
\label{tab:parafp}
\begin{threeparttable}
\begin{tabular}{cclll}
\toprule[1pt]
\midrule
 & $\left(\mathrm{sig.}, \mathrm{exp.}\right)$\tnote{(1)} & \multicolumn{1}{c}{$u$\tnote{(2)}} & \multicolumn{1}{c}{$x_{\min}$\tnote{(3)}} & \multicolumn{1}{c}{$x_{\max}$\tnote{(4)}} \\ \midrule
$\mathtt{bfloat16}$    &  $(8,8)$    & $3.91\times 10^{-3}$ & $1.18\times 10^{-38}$& $3.39\times 10^{38}$ \\ 
$\mathtt{fp16}$      &  $(11,5)$    & $4.88\times 10^{-4}$ & $6.10\times 10^{-5}$& $6.55\times 10^{4}$ \\ 
$\mathtt{fp32}$     &  $(24,8)$    & $5.96\times 10^{-8}$ & $1.18\times 10^{-38}$& $3.40\times 10^{38}$ \\ 
$\mathtt{fp64}$     &  $(53,11)$    & $4.88\times 10^{-16}$ & $2.22\times 10^{-308}$& $1.80\times 10^{308}$ \\ \midrule
\bottomrule[1pt]
\end{tabular}
    \begin{tablenotes}    
        \footnotesize               
        \item[(1)] $\left(\mathrm{sig.}, \mathrm{exp.}\right)$ represents number of bits in significand and exponent.          
        \item[(2)] $u=\frac{1}{2}\eta^{1-t}$ is unit roundoff.
        \item[(3)] $x_{\min}$ is smallest normalized positive number.
        \item[(4)] $x_{\max}$ is largest finite number.
    \end{tablenotes}            
\end{threeparttable}
\end{table} 

\begin{assumption}[\textit{Rounding errors modeling}]
\label{ass:rv_error}
    Following \cite{higham2019new} and \cite{connolly2021stochastic}, the quantities $\delta$ in the model \ref{def:samodel} associated with every pair of operands are modeled as independent random variables of mean zero.
\end{assumption}

% Analyzing rounding errors poses a considerable challenge due to the absence of an exact distribution for these errors, which motivates us to find the closed-form bounds of rounding errors. 
Under \textit{Assumption} \ref{ass:rv_error}, we can give the error bound for the real-valued matrix-matrix products with finite-precision arithmetic in the following lemma.

\begin{lemma}[\textit{Real-valued matrix-matrix products}{\cite[Theorem 3.4]{higham2019new}}] 
\label{lem:rv_inner}
    Let $\mathbf{C}=\mathbf{A}{\bf B}$, where $\mathbf{A} \in \mathbb{R}^{m\times n},\mathbf{B}\in \mathbb{R}^{n\times p}$, be evaluated in the finite-precision arithmetic. Under \textit{Assumption} \ref{ass:rv_error}, the computed $\mathbf{C}^{(l)}$ satisfies
    \begin{equation}
    \label{eq:rv_bound_inner}
        \left| \mathbf{C}^{(l)} - \mathbf{C}\right| \leq \gamma_n \left| \mathbf{A}\right|\left| \mathbf{B}\right|,
    \end{equation}
    where the superscript $(l)$ denotes the corresponding result of computation using finite-precision arithmetic, and we define $\gamma_n$ by
    \begin{equation}
    \label{eq:gamma}
        \gamma_n = \exp{\left(\lambda\sqrt{n}u+\frac{nu^2}{1-u}\right)} -1,
    \end{equation}
    where { $\lambda$ is a positive constant that can be freely chosen and controls the probability of failure of the bound, which is a monotonically decreasing function of $\lambda$.} Please see the references for the details \cite{higham2019new,connolly2021stochastic, doi:10.1137/20M1314355}. 
\end{lemma} 
%Probabilities of failure are not explicitly stated, as it is close to 1 even for modest $\lambda$ and leaning toward pessimism.
% \textit{Lemma} \ref{lem:rv_inner} clearly illustrates that the presence of dimension $n$ in bound \eqref{eq:rv_bound_inner}, signifying the accumulation of rounding errors along the dimension of the vector, may impede computation accuracy when $n$ or $u$ is larger, i.e., high vector dimension or low-precision arithmetic. 
% \begin{remark}[\textit{Impact of $\lambda$}]
% \label{rem:lambda}
%     $\lambda$ is a positive constant that can be freely chosen and controls the probability of failure of the bound, which is a monotonically decreasing function of $\lambda$. Probabilities of failure are not explicitly stated, as it is close to 1 even for modest $\lambda$ and leaning toward pessimism. Please see the references for the details \cite{higham2019new,connolly2021stochastic,doi:10.1137/20M1314355}.
% \end{remark}

\section{Single-User Scenario with Finite-precision Arithmetic}
\label{sec:su}
In this section, we focus on the special case of a single user when $K=1$, i.e., SIMO and MISO systems, as shown in Fig. \ref{fig:su}. In this context, the linear detection matrix and precoding matrix simplify to MRC vector ${\bf a}$ and MRT vector ${\bf p}$, respectively. 

Note that the channel matrix $\bf H$ and other related parameters are modeled as complex matrices and vectors in Sec. \ref{sec:sys_de}. Consequently, \textit{Lemma} \ref{lem:rv_inner} is not directly applicable in this context. To the best of our knowledge, the rounding error of complex-valued matrices and vectors has not been derived before. Hence, we extend the error bounds in \textit{Lemma} \ref{lem:rv_inner} to encompass complex-valued arithmetic in the theorem below.
% Note that communication systems always employ complex-valued arithmetic. However, the rounding error bounds of existing literature cater exclusively to real-valued arithmetic.

\begin{theorem}[\textit{Complex-valued inner products}]
\label{the:inner products}
    Let ${ s}=\mathbf{a}^{H}\mathbf{b}$, where $\mathbf{a,b}\in \mathbb{C}^{n\times 1}$, be evaluated in the finite-precision arithmetic. Under \textit{Assumption} \ref{ass:rv_error}, the computed ${s}^{(l)}$ satisfies
    \begin{equation}
    \label{eq:bound_inner}
        \left\| {s}^{(l)} -{s} \right\|_2 \leq \sqrt{2} \gamma_{2n} \left\| \mathbf{a} \right\|_2\left\| \mathbf{b} \right\|_2.
    \end{equation}
\end{theorem} %if all operations are carried out in a uniform precision $u$,
\begin{IEEEproof}
    The proof is available in Appendix \ref{app:inner products}.
\end{IEEEproof}
\textit{Theorem} \ref{the:inner products} reflects the fact that rounding errors accumulate along the vector dimension, which may prevent the computation from achieving sufficient accuracy when employing low-precision arithmetic or increasing the dimension. Moreover, \textit{Theorem} \ref{the:inner products} can be extended to encompass complex-valued matrix-vector and matrix-matrix products as follows:
%seen in the Appendix \ref{app:extend_the1}
\begin{theorem}[\textit{Complex-valued matrix-vector and matrix-matrix products}]
 \label{the:cv_mv_mm}   
    Let $\mathbf{A}\in \mathbb{C}^{m\times n}$, $\mathbf{B}\in \mathbb{C}^{n\times p}$, and $\mathbf{x}\in \mathbb{C}^{n\times 1}$. Under \textit{Assumption} \ref{ass:rv_error}, if ${\bf y = Ax}$ is carried out in the finite-precision arithmetic, the computed $\mathbf{y}^{(l)}$ satisfies 
    \begin{equation}
        \left\| \mathbf{y}^{(l)} -\mathbf{y} \right\|_2 \leq \sqrt{2\min{(m,n)}} \gamma_{2n} \left\| \mathbf{A} \right\|_2\left\| \mathbf{x} \right\|_2.
    \end{equation}
    Under \textit{Assumption} \ref{ass:rv_error}, if ${\bf C = AB}$ is carried out in the finite-precision arithmetic, the computed $\mathbf{C}^{(l)}$ satisfies  
    \begin{equation}
    \small
    \label{eq:bound_mmp}
        \left\| \mathbf{C}^{(l)} -\mathbf{C} \right\|_2 \leq 2\sqrt{\min{(m,n)}\min{(n,p)}} \gamma_{2n} \left\| \mathbf{A} \right\|_2\left\| \mathbf{B} \right\|_2.
    \end{equation}
\end{theorem}
\begin{IEEEproof}
    The proof is similar to that of \textit{Theorem} \ref{the:inner products}, which is omitted for conciseness.
\end{IEEEproof}

% First, we derive the lower bound for the ergodic achievable rate in SIMO and MISO systems, respectively. Then a comparative analysis between SIMO and MISO systems is conducted.

\subsection{SIMO Using MRC with Finite-Precision Arithmetic}
\begin{figure}[t]
  \centering
    \subfloat[Uplink SIMO with finite-precision arithmetic]{\includegraphics[width=0.3\textwidth]{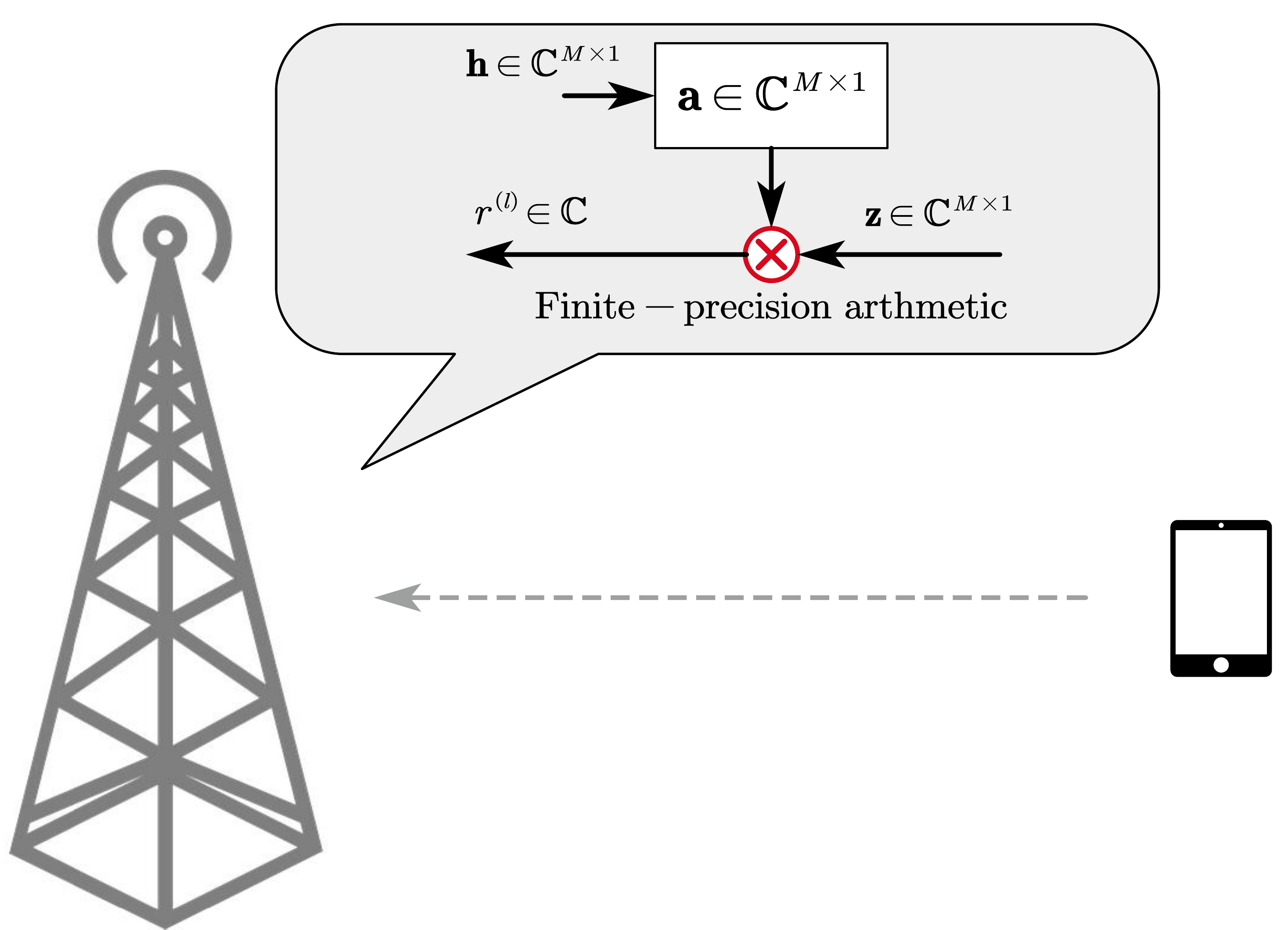}
    \label{fig:simo}}
    \hfill
    \subfloat[Downlink MISO with finite-precision arithmetic]{\includegraphics[width=0.3\textwidth]{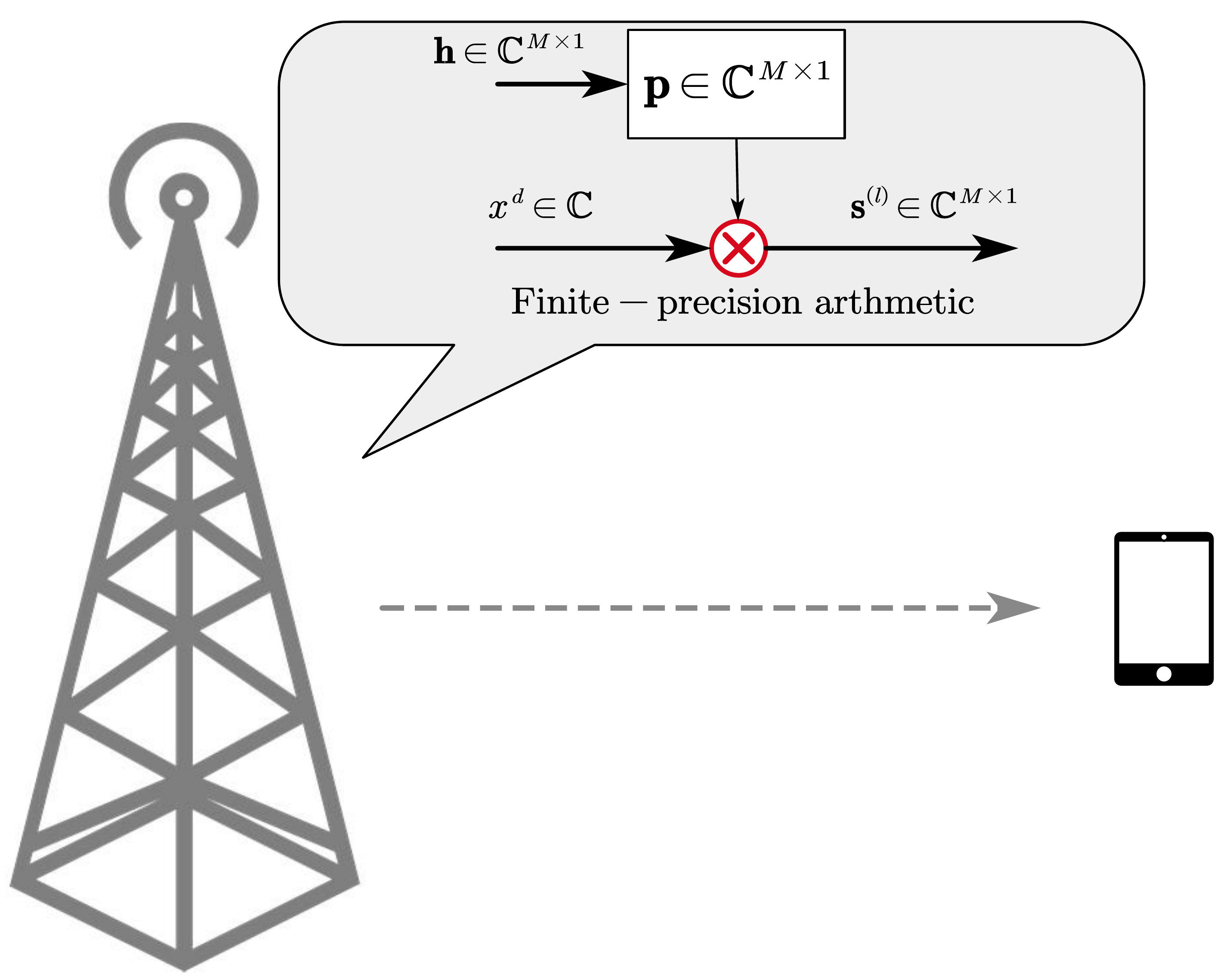}
    \label{fig:miso}}
    \caption{The illustration of single-user scenario with finite-precision arithmetic.}
    \label{fig:su}
\end{figure}

As is shown in Fig. \ref{fig:simo}, if the process of combining is carried out in the finite-precision arithmetic, we can derive the rounding error bound in the following lemma.
\begin{lemma}[\textit{Error bound for SIMO}]
 \label{lem:simo}
    When \eqref{eq:up_linear} is carried out in the finite-precision arithmetic, the received signal after MRC at the BS can be expressed as
    \begin{equation}
        \begin{aligned}
                r^{(l)} & = r + \Delta r \\
               & =  \sqrt{\rho}{\bf a}^{H}{\bf h}{x}^\mathrm{u} + {\bf a}^{H}{\bf n} + \Delta r \\
               & = \sqrt{\rho}{\bf h}^{H}{\bf h}{x}^\mathrm{u} + {\bf h}^{H}{\bf n} + \Delta r,
        \end{aligned}
    \end{equation}
    where $\Delta r \in \mathbb{C}$ satisfies
    \begin{equation}
    \label{eq:simo_r}
        \left\| \Delta r \right\|_2 \leq \delta_{\mathrm{SIMO}} \left\| \mathbf{h} \right\|_2\left\| \mathbf{z} \right\|_2,
    \end{equation}
    where $\delta_{\mathrm{SIMO}} = \sqrt{2} \gamma_{2M}$.   
\end{lemma}
\begin{IEEEproof}
     The proof follows directly from \textit{Theorem} \ref{the:inner products} by defining $\Delta r=r^{(l)}-r$.
\end{IEEEproof}
\textit{Lemma} \ref{lem:simo} shows that except for the impact of low-precision arithmetic, rounding errors accumulate with the increasing number of antennas $M$. Moreover, under \textit{Assumption} \ref{ass:rv_error}, rounding error $\Delta r$ is a random variable. By assuming the noise is independent of rounding error, the ergodic achievable rate of the SIMO system is
% Therefore, in massive MIMO systems or extremely large MIMO systems,  \cite{7886292,6816003}
\begin{equation}
\label{eq:er_simo}
    R_{\mathrm{SIMO}} = \mathbb{E}\left\{ \log_2 \left(1+\frac{\rho\left\| \mathbf{h} \right\|_2^4}{\left\| \mathbf{h} \right\|_2^2+\left\| \Delta r \right\|_2^2} \right)\right\}.
\end{equation}

Note that deriving an approximate closed-form expression of $\eqref{eq:er_simo}$ remains challenging due to the absence of an exact distribution for rounding errors, which motivates us to find the closed-form lower bound of \eqref{eq:er_simo} as shown in the proposition below.
\begin{proposition}[\textit{Lower bound of the achievable rate for SIMO}]
\label{prop:lower_bound_simo}
    Using MRC with finite-precision arithmetic in Rayleigh fading, the achievable rate for SIMO can be lower bounded as follows:
    \begin{equation}
    \label{eq:lower_bound_simo}
    \breve{R}_{\mathrm{SIMO}} = \log_2 \left(1+\frac{\rho M}{1+\delta_{\mathrm{SIMO}}^2M(\rho + 1)} \right).
    \end{equation}
\end{proposition}
\begin{IEEEproof}
    Substituting \eqref{eq:simo_r} into \eqref{eq:er_simo}, we obtain
        \begin{align}
            {R}_{\mathrm{SIMO}}&\ge \mathbb{E} \left\{ \log _2\left( 1+\frac{\rho \left\| \mathbf{h} \right\| _{2}^{4}}{\left\| \mathbf{h} \right\| _{2}^{2}+\delta _{\mathrm{SIMO}}^{2}\left\| \mathbf{h} \right\| _{2}^{2}\left\| \mathbf{z} \right\| _{2}^{2}} \right) \right\}\\
	&=\mathbb{E} \left\{ \log _2\left( 1+\frac{\rho \left\| \mathbf{h} \right\| _{2}^{2}}{1+\delta _{\mathrm{SIMO}}^{2}\left\|  \mathbf{z} \right\| _{2}^{2}} \right) \right\}\\
	&\overset{\left( a \right)}{\approx}\log _2\left( 1+\frac{\rho\mathbb{E} \left\{  \left\| \mathbf{h} \right\| _{2}^{2} \right\}}{1+\delta _{\mathrm{SIMO}}^{2}\mathbb{E} \left\{ \left\| \mathbf{z} \right\| _{2}^{2} \right\}} \right)\\
	&\overset{\left( b \right)}{=}\log _2\left( 1+\frac{\rho M}{1+\delta _{\mathrm{SIMO}}^{2}M(\rho +1)} \right),
        \end{align}
    where we employ \cite[lemma 1]{6816003} at the point $(a)$, and we have $\mathbb{E}\left\{  \left\| \mathbf{h} \right\| _{2}^{2} \right\} = M$ and $\mathbb{E}\left\{  \left\|  \mathbf{z} \right\| _{2}^{2} \right\} =\mathbb{E} \left\{ \rho (x^\mathrm{u})^H\mathbf{h}^H\mathbf{h}x^\mathrm{u}+\mathbf{n}^H\mathbf{n} \right\} =\rho M+M=M\left( \rho +1 \right) $ at the point $(b)$.
\end{IEEEproof}
\textit{Proposition} \ref{prop:lower_bound_simo} reveals that the impact of the SNR $\rho$, the number of BS antennas $M$ and the precision $u$. It is clear that if we use full-precision arithmetic, i.e., $u\rightarrow 0$, \eqref{eq:lower_bound_simo} reduces to the special case of Rayleigh fading channel \cite[Eq. (21)]{lim2015performance}. We further show the relationship between the achievable rate in \textit{Proposition} \ref{prop:lower_bound_simo} and key parameters by providing the following important analysis.
\begin{corollary}[\textit{Impact of $M$ for SIMO}]
    \label{coro:impact_M_simo}
    For a fixed $\rho$ and fixed precision arithmetic, \eqref{eq:lower_bound_simo} is a first increasing and then decreasing function of $M$, and the global maximum point $M_{\max}$ can be expressed as 
    \begin{equation}
    \label{eq:M_max}
        M_{\max} \approx \left \lfloor \frac{1}{2u\lambda \sqrt{\rho +1}} \right \rfloor.
    \end{equation}
    More specifically, when $M$ grows without bound, \eqref{eq:lower_bound_simo} tends to
    \begin{equation}
    \label{eq:M_infity}
        \lim_{M \rightarrow \infty} \breve{R}_{\mathrm{SIMO}}= 0.
    \end{equation}
\end{corollary}
\begin{IEEEproof}
    For simplicity, we use the Taylor expressions on the \eqref{eq:gamma} and have %\cite{higham2019new}
    \begin{equation}
        \gamma_n = \exp{\left(\lambda\sqrt{n}u+\frac{nu^2}{1-u}\right)} -1 = \lambda\sqrt{n}u + O(u^2).
    \end{equation}
    Then we further ignore the high-order term and obtain $\gamma_n\approx \lambda\sqrt{n}u$. For $n =1000$, $\lambda=1$ and using $\mathtt{fp16}$ arithmetic, i.e., $u=4.88\times 10^{-4}$, this approximation error is only $3.62\times 10^{-4}$. Plugging the approximation expression into \eqref{eq:lower_bound_simo}, we have
    \begin{equation}
    \label{eq:approma_lower_bound_simo}
        \breve{R}_{\mathrm{SIMO}} \approx \log_2 \left(1+\frac{\rho M}{1+4u^2\lambda ^2\left(\rho+1 \right)M^2} \right).
    \end{equation}
    Note that $\log_2(1+x)$ is a monotonically increasing function of $x$. Hence, we only need to analyze the monotonicity of $f(M) = \frac{\rho M}{1+4u^2\lambda ^2\left(\rho+1 \right)M^2}$. Then the first-order and second-order partial derivatives of $f(M)$ in terms of $M$ can be calculated as
    {\small\begin{align}
        \frac{\partial f(M)}{\partial M} & = \frac{\rho\left( 1-4M^2u^2\lambda ^2\left( 1+\rho \right) \right)}{\left( 1+4M^2u^2\lambda ^2\left( 1+\rho \right) \right) ^2}, \label{eq:first-order}\\
        \frac{\partial ^2f(M)}{\partial M^2} & =\frac{8\rho\left( 1+\rho \right) u^2\lambda ^2M\left( 4M^2u^2\lambda ^2\left( 1+\rho \right) -3 \right)}{\left( 1+4M^2u^2\lambda ^2\left( 1+\rho\right) \right) ^3}. \label{eq:second-order}
    \end{align}}
    Utilizing first and second-order derivative conditions, we can readily ascertain that the local maximum point is the global maximum point $M_{\max}$, and \eqref{eq:lower_bound_simo} is a first increasing and then decreasing function of $M$.
    
    At last, let $M$ grow without bound, we can directly observe that $\breve{R}_{\mathrm{SIMO}} \rightarrow 0$ from \eqref{eq:approma_lower_bound_simo}.
\end{IEEEproof}
It is interesting to note from \textit{Corollary} \ref{coro:impact_M_simo} that when considering finite-precision arithmetic, the lower bound of the rate\footnote{Although we only derive the lower bound on the achievable rate, simulations in Sec. \ref{sec:sim} will affirm that this lower bound effectively mirrors the changing trends of the achievable rate.} will not exhibit a monotonic increase with the number of BS antennas; instead, it follows an initial increase and subsequent decrease. Leveraging optimization theory, we affirm the existence of an optimal number of BS antennas that maximizes the lower bound of the rate. Moreover, the value provided in \eqref{eq:M_max} may not be an integer, necessitating rounding to the nearest integer to determine the optimal number of BS antennas.
%This is reasonable since rounding errors will accumulate as the number of BS antennas grows.

\begin{corollary}[\textit{Impact of $\rho$ for SIMO}]
    \label{coro:impact_rho_simo}
    For a fixed $M$ and fixed precision arithmetic, when the SNR goes without bound, \eqref{eq:lower_bound_simo} tends to
    \begin{equation}
    \label{eq:rho_infity}
        \lim_{\rho \rightarrow \infty} \breve{R}_{\mathrm{SIMO}}=\log _2\left( 1+\delta _{\mathrm{SIMO}}^{-2} \right).
    \end{equation}
\end{corollary}
\begin{IEEEproof}
    The desired result can be directly derived by letting $\rho \rightarrow \infty$ in \eqref{eq:lower_bound_simo}.
\end{IEEEproof} 
\textit{Corollary} \ref{coro:impact_rho_simo} indicates that due to the impact of finite-precision arithmetic, the lower bound of the rate will approach a ceiling as $\rho$ grows. 

\subsection{MISO Using MRT with Finite-Precision Arithmetic}
As illustrated in Fig. \ref{fig:miso}, if the process of detection is carried out in the finite-precision arithmetic, we can obtain the rounding error bound in the following lemma.
\begin{lemma}[\textit{Error bound for MISO}]
    \label{lem:miso}
    When \eqref{eq:precoding} is carried out in the finite-precision arithmetic, the transmit vector after MRT at the BS can be expressed as
    \begin{equation}
    \label{eq:miso_s_d}
        \begin{aligned}
                {\bf s}^{(l)} & = {\bf s} + \Delta {\bf s}  \\
               & =  \sqrt{\beta}{\bf p}{ x}^{\mathrm{d}} + \Delta {\bf s} \\
               & = \frac{{\bf h}}{\left\| \mathbf{h} \right\|_2}{ x}^{\mathrm{d}} + \Delta {\bf s}, 
        \end{aligned}
    \end{equation}
    where $\Delta {\bf s} \in \mathbb{C}^{M\times 1}$ satisfies
    \begin{equation}
    \label{eq:miso_s}
        \left\| \Delta {\bf s} \right\|_2 \leq  \frac{{1}}{\left\| \mathbf{h} \right\|_2} \delta_{\mathrm{MISO}}\left\| \mathbf{h} \right\|_2 \left\| x^\mathrm{d} \right\|_2 = \delta_{\mathrm{MISO}}\left\| x^\mathrm{d} \right\|_2,
    \end{equation}
    where $\delta_{\mathrm{MISO}} = \sqrt{2} \gamma_{2}$.
\end{lemma}
\begin{IEEEproof}
     The proof follows directly from \textit{Theorem} \ref{the:cv_mv_mm} by substituting $\bf A$ and $\bf x$ with $\bf p$ and $x^\mathrm{d}$, respectively.
\end{IEEEproof}
Compared between \textit{Lemma} \ref{lem:miso} and \textit{Lemma} \ref{lem:simo}, it can be observed that rounding errors are fixed and independent of the number of antennas $M$ in MISO systems. This is because of its only involvement in \textit{vector-scalar products}, i.e., a series of finite-precision \textit{scalar-scalar products}, for MISO systems. Moreover, the rounding errors in MISO systems are not influenced by noise, as MRT is implemented at the transmitter.
% Conversely, for SIMO systems, it is involved in \textit{inner products}, i.e., a series of finite-precision scalar-scalar \textit{products} and \textit{summation}, leading to cumulative effects as $M$ increases.

Furthermore, by substituting \eqref{eq:dl_linear_mu} with \eqref{eq:miso_s_d}, we can obtain
\begin{equation}
    \begin{aligned}
        y^{(l)} & = \sqrt{\rho}\mathbf{h}^H {\bf s}^{(l)} \\
    & = \sqrt{\rho}\left\| \mathbf{h} \right\| x^\mathrm{d} + \sqrt{\rho}\mathbf{h}^H \Delta {\bf s} + n. 
    \end{aligned}
\end{equation}

Similar to the analysis of SIMO, the ergodic achievable rate of the MISO system is
\begin{equation}
\label{eq:er_miso}
    R_{\mathrm{MISO}} = \mathbb{E} \left\{ \log _2\left( 1+\frac{\rho \left\| \mathbf{h} \right\|_2 ^2}{\rho \left| \mathbf{h}^H\Delta{\bf s} \right|^2+1} \right) \right\}.
\end{equation}

Next, we will investigate the impact of finite-precision arithmetic for MISO by \eqref{eq:er_miso}; this is done by first presenting the lower bound on the achievable rate in the proposition below.

\begin{proposition}[\textit{Lower bound of the achievable rate for MISO}]
\label{prop:lower_bound_miso}
    Using MRT with finite-precision arithmetic in Rayleigh fading, the achievable rate for MISO can be lower bounded as
    \begin{equation}
    \label{eq:lower_bound_miso}
    \breve{R}_{\mathrm{MISO}} = \log_2 \left(1+\frac{\rho M}{1+\delta_{\mathrm{MISO}}^2\rho M} \right).
    \end{equation}
\end{proposition}
\begin{IEEEproof}
    Using Cauchy-Schwartz inequity, \eqref{eq:er_miso} can be expressed as
        \begin{align}
            R_{\mathrm{MISO}} & \geq \mathbb{E} \left\{ \log _2\left( 1+\frac{\rho \left\| \mathbf{h} \right\|_2^2}{\rho\left\| \mathbf{h} \right\|_2 ^2 \left\| \Delta{\bf s} \right\|_2^2+1} \right) \right\} \\
            &\overset{\eqref{eq:miso_s}}{\geq} \mathbb{E} \left\{ \log _2\left( 1+\frac{\rho \left\| \mathbf{h} \right\|_2^2}{\rho\left\| \mathbf{h} \right\|_2^2 \delta_{\mathrm{MISO}}^2\left\| x^\mathrm{d} \right\|_2^2 +1} \right) \right\} \\
            &\overset{\left( c \right)}{\approx} \log _2\left( 1+\frac{\rho \mathbb{E} \left\{ \left\| \mathbf{h} \right\|_2^2\right\}}{\delta_{\mathrm{MISO}}^2 \rho \mathbb{E} \left\{ \left\| \mathbf{h} \right\|_2^2 \left\| x^\mathrm{d} \right\|_2^2\right\} +1} \right) \\
           & =  \log_2 \left(1+\frac{\rho M}{1+\delta_{\mathrm{MISO}}^2\rho M} \right),
        \end{align}
    where we employ \cite[lemma 1]{6816003} at the point $(c)$.
\end{IEEEproof}
Similar to \textit{Proposition} \ref{prop:lower_bound_simo}, \textit{Proposition} \ref{prop:lower_bound_miso} elucidates the influence of three crucial factors: the SNR $\rho$, the number of BS antennas $M$, and the precision parameter $u$. It is evident that under full-precision arithmetic, where $u$ tends to 0, \eqref{eq:lower_bound_miso} reduces to the specific scenario of a Rayleigh fading channel \cite[Eq. (11)]{lim2015performance}. We further show the relationship between the achievable rate in \textit{Proposition} \ref{prop:lower_bound_miso} and key factors through the subsequent analysis.

\begin{corollary}[\textit{Impact of $M$ for MISO}]
    \label{coro:M_miso}
    For a fixed $\rho$ and fixed precision arithmetic, when $M$ grows without bound, \eqref{eq:lower_bound_miso} tends to
    \begin{equation}
         \lim_{M \rightarrow \infty} \breve{R}_{\mathrm{MISO}} = \log _2\left( 1+\delta _{\mathrm{MISO}}^{-2} \right).
    \end{equation}
\end{corollary}
\begin{IEEEproof}
    The desired result can be directly derived by letting $M \rightarrow \infty$ in \eqref{eq:lower_bound_miso}.
\end{IEEEproof} 
\begin{corollary}[\textit{Impact of $\rho$ for MISO}]
    \label{coro:rho_miso}
    For a fixed $M$ and fixed precision arithmetic, when $\rho$ grows without bound, \eqref{eq:lower_bound_miso} converges to
    \begin{equation}
         \lim_{\rho \rightarrow \infty} \breve{R}_{\mathrm{MISO}} = \log _2\left( 1+\delta _{\mathrm{MISO}}^{-2} \right).
    \end{equation}
\end{corollary}
\begin{IEEEproof}
    The desired result can be directly derived by letting $\rho \rightarrow \infty$ in \eqref{eq:lower_bound_miso}.
\end{IEEEproof} 
Similar to \textit{corollary} \ref{coro:impact_rho_simo}, both \textit{corollary} \ref{coro:M_miso} and \ref{coro:rho_miso} reveal that as $M$ or $\rho$ grows, the lower bound of the rate will converge to an exact rate. 

\subsection{Insights for SIMO and MISO with Finite-Precision Arithmetic}
Based on the derivation and analysis presented in the preceding two subsections, we can summarize several noteworthy insights regarding SIMO and MISO systems employing finite-precision arithmetic.

\begin{remark}[\textit{Increasing the BS antennas is not always beneficial for SIMO}]
    From \textit{Corollary} \ref{coro:impact_M_simo}, we find that the lower bound of the rate for SIMO systems does not show a monotonic increase with the BS antennas $M$, and tends to zero as $M$ goes to infinity. This is reasonable since rounding errors will accumulate as the number of BS antennas grows.
\end{remark}

\begin{remark}[\textit{Floor effect of the rounding errors for SIMO and MISO}]
\label{rem:floor_effect_simo}
    We can observe that as the SNR $\rho$ increases, the lower bound of the rate for SIMO systems approaches a constant value. This phenomenon arises from the fixed nature of rounding errors for a given $M$, where the effect of finite-precision arithmetic manifests as an additive noise. Similarly, for MISO systems, the lower bound of the rate also converges to a fixed rate with the increase of $M$ or $\rho$. This is attributed to the independence of the rounding error in MISO systems with respect to $M$ and $\rho$, treating it as a fixed additive noise.
\end{remark}

% \begin{remark}[\textit{Floor effect of rounding error for MISO}]
%     Similar to \textit{Remark} \ref{rem:floor_effect_simo}, 
% \end{remark}

\begin{remark}[\textit{Duality between SIMO and MISO with finite-precision arithmetic fails}]
    Duality between SIMO and MISO, i.e., $R_{\mathrm{MISO}} = R_{\mathrm{SIMO}}$, is widely acknowledged with full-precision arithmetic \cite{1237143}. However, as revealed in the preceding analysis, it is interesting to note that there exists a performance gap between SIMO and MISO systems with finite-precision arithmetic. This is because they utilize different finite-precision arithmetic. Specifically, for SIMO systems, it is involved in \textit{inner products}, i.e., a series of finite-precision \textit{scalar-scalar products} and \textit{summation}, leading to cumulative effects as $M$ increases. In contrast, MISO systems use \textit{vector-scalar products}, i.e., a series of finite-precision \textit{scalar-scalar products}, which remain unaffected by the growth of $M$.
\end{remark}

\begin{remark}[\textit{Performance gap between SIMO and MISO}]
    Let us denote the rate gap $\Delta R$ as follows:
    \begin{equation}
    \label{eq:delta_R}
        \Delta R = \breve{R}_{\mathrm{MISO}} - \breve{R}_{\mathrm{SIMO}}.
    \end{equation}
    For a fixed $\rho$ and fixed precision arithmetic, when $M$ grows without bound, \eqref{eq:delta_R} tends to
    \begin{equation}
         \lim_{M \rightarrow \infty} \Delta R= \log _2\left( 1+\delta _{\mathrm{MISO}}^{-2} \right).
    \end{equation}
    For a fixed $M$ and fixed precision arithmetic, when $\rho$ grows without bound, \eqref{eq:delta_R} tends to
    \begin{equation}
         \lim_{\rho \rightarrow \infty} \Delta R= \log _2\left( 1+\delta _{\mathrm{MISO}}^{-2} \right) - \log _2\left( 1+\delta _{\mathrm{SIMO}}^{-2} \right).
    \end{equation}
\end{remark}

\section{General Multi-User Scenario with Finite-Precision Arithmetic}
\label{sec:mu}
This section mainly investigates the performance under the general multi-user setup, i.e., MU-SIMO and MU-MISO systems, with finite-precision arithmetic. To avoid inter-user interference, we use ZF detection in the uplink and ZF precoding in the downlink, respectively. First, we will derive the rounding error analysis and lower bound on the achievable sum rate for MU-SIMO systems. And then the lower bound on the achievable sum rate for MU-MISO systems will be presented.

\subsection{MU-SIMO Using ZF Detection with Finite-Precision Arithmetic}
We first consider an uplink scenario. From the previous discussion, BS utilizes ZF detection to avoid inter-user interference, i.e., 
\begin{equation}
    \label{eq:zfd}
    {\bf A} =  {\bf H}\left ({\bf H}^H{\bf H} \right)^{-1}.
\end{equation}
Then the received vector after using the linear detector is given by
\begin{equation}
\label{eq:zf_c}
    {\bf r} = \left ({\bf H}^H{\bf H} \right)^{-1}{\bf H}^H{\bf z}.
\end{equation}

%First, the computation of ZF detection, i.e., \eqref{eq:zf}, necessitates finite-precision arithmetic. Second, the combining process, i.e., \eqref{eq:zf_c}, should also be conducted in the finite-precision arithmetic.
Note that, different from the analysis for SIMO systems, rounding error analysis in MU-SIMO systems detection is challenging due to matrix inversion in \eqref{eq:zfd} or \eqref{eq:zf_c}. To avoid matrix inversion, we transform \eqref{eq:zf_c} into LS problem, yielding
\begin{equation}
\label{eq:ne}
    {\bf H}^H{\bf H} {\bf r}^{(l)} = {\bf H}^H{\bf z}, 
\end{equation}
where ${\bf r}^{(l)}$ is the received vector after detection with finite-precision arithmetic and each computing is carried out in the finite-precision arithmetic. 

One traditional approach for solving the LS problem is the NE method\footnote{Compared to the iterative method, the NE method is more convenient for rounding error analysis because it relies on standard algorithms such as Cholesky factorization, matrix-matrix multiplication, and matrix-vector multiplication.} \cite[Alg. 5.3.1]{gloub1996matrix}. In this context, we present the NE method-based ZF detection, as outlined in \textit{Algorithm} \ref{alg:ne_simo}. Notably, in the massive MIMO systems where $M\gg K$, the NE method requires only half flops of the QR factorization approach. Furthermore, we can derive the rounding error bound in the lemma below.

\begin{algorithm}[t]
    \caption{NE method-based ZF detection for MU-SIMO systems with finite-precision arithmetic}%标题
    \label{alg:ne_simo}
      \LinesNumbered
       \KwIn {Channel matrix $\mathbf{H}$, received vector $\mathbf{z}$.}
      \KwOut {${\bf r}^{(l)}$.}
      Compute the matrix-vector products 
        \begin{equation}\setlength\abovedisplayskip{1.5pt}
            \setlength\belowdisplayskip{1.5pt}
            {\bf c}^{(l)}=\boldsymbol{fl}\left(\mathbf{H}^H\mathbf{z}\right). \notag
        \end{equation}
        
        Compute the matrix-matrix products 
        \begin{equation}\setlength\abovedisplayskip{1.5pt}
            \setlength\belowdisplayskip{1.5pt}
            {\bf C}^{(l)}=\boldsymbol{fl}\left(\mathbf{H}^H\mathbf{H}\right). \notag
        \end{equation}

        Compute the Cholesky factorization 
        \begin{equation}\setlength\abovedisplayskip{1.5pt}
            \setlength\belowdisplayskip{1.5pt}
            {\bf C}^{(l)}=({\bf R}^{(l)})^H{\bf R}^{(l)}. \notag
        \end{equation}
 % based on \cite[Alg. 4.2.2]{gloub1996matrix}

        Solve the triangular equations
        \begin{equation}\setlength\abovedisplayskip{1.5pt}
            \setlength\belowdisplayskip{1.5pt}
            \begin{array}{cc}
              ({\bf R}^{(l)})^H{\bf q}^{(l)}={\bf c}^{(l)},   &  {\bf R}^{(l)}{\bf r}^{(l)}={\bf q}^{(l)}. \notag
            \end{array}
        \end{equation}
   % based on \cite[Alg. 3.1.1 \& 3.1.2]{gloub1996matrix}
\end{algorithm}

\begin{lemma}[\textit{Error bound for MU-SIMO}]
\label{lem:MU-SIMO}
    When ZF detection is carried out in the finite-precision arithmetic based on \textit{Algorithm} \ref{alg:ne_simo}, and provided that $c_1^{\mathrm{u}}\kappa _2\left( \mathbf{H}^H\mathbf{H} \right)<1$ while ensuring the backward stability of \textit{Algorithm} \ref{alg:ne_simo}, the received signals after ZF detection at the BS can be expressed as 
    \begin{equation}
        {\bf r}^{(l)}  = {\bf r} + \Delta{\bf r}  = {\bf A}^{H}{\bf z} + \Delta{\bf r}, 
    \end{equation}
    where $\Delta{\bf r} \in \mathbb{C}^{K\times 1}$ satisfies
    \begin{equation}
    \label{eq:ne_cv}
        \left\| \Delta{\bf r} \right\| _2\leq c^{\mathrm{u}}\kappa _2\left( \mathbf{H}^H\mathbf{H} \right) \left\| \mathbf{r} \right\| _2,
    \end{equation}
    with 
    \begin{equation}\small\label{eq:c_musimo}
    \begin{aligned}
                c^{\mathrm{u}} &= c_1^{\mathrm{u}} +\sqrt{2K}\gamma _{2M}, \\
        c_1^{\mathrm{u}}  &= 2K\left( \gamma _{2M}+\gamma _{6K+1}/\left( 1-2K\gamma _{2K+1} \right)\right),
    \end{aligned}
    \end{equation}
    where $K$ is the number of users. And the received signal for the $k$th user is given by
    \begin{equation}
    \begin{aligned} 
        { r}^{(l)}_k &= \sqrt{\rho}{x}_k^\mathrm{u} + \sqrt{\rho}\sum_{i=1,i\neq k}^{K}{\bf a}_k^{H}{\bf h}_i{x}_i^\mathrm{u} +{\bf a}_k^{H}{\bf n} + \Delta{\bf r}_k \\
        & = \sqrt{\rho}{x}_k^\mathrm{u} + {\bf a}_k^{H}{\bf n} + \Delta{\bf r}_k,
    \end{aligned}        
    \end{equation}
\end{lemma}
\begin{IEEEproof}
    The proof is available in Appendix \ref{app:NE_D}.
\end{IEEEproof}
\textit{Lemma} \ref{lem:MU-SIMO} demonstrates that rounding errors accumulate as the numbers of antennas $M$ and users $K$ increase. Additionally, the condition number of the channel matrix plays a crucial role in influencing the rounding errors. This is reasonable since \textit{Algorithm} \ref{alg:ne_simo} is involved in the Cholesky factorization and solving the triangular equations, making it more demanding in terms of channel condition compared with matrix-matrix products. Moreover, by neglecting the rounding error in \textit{Steps} 1 and 2 of \textit{Algorithm} \ref{alg:ne_simo}, \eqref{eq:c_musimo} reduces to $c^{\mathrm{u}} = 2K\gamma _{6K+1}/\left( 1-2K\gamma _{2K+1}\right)$, aligning with the specific case presented in \cite[Eq. (5.3.2)]{gloub1996matrix}.

Based on \textit{Lemma} \ref{lem:MU-SIMO}, the ergodic achievable sum rate of the MU-SIMO system can be expressed as
\begin{equation}\small
\label{eq:rate_ms}
    R_{\mathrm{MS}} = \sum_{k=1}^{K}\mathbb{E} \left\{ \log _2\left( 1+\frac{\rho}{\left[ \left( \mathbf{H}^H\mathbf{H} \right) ^{-1} \right] _{kk}+\left\| \Delta{\bf r}_k \right\| _2^2} \right) \right\}.
\end{equation}
Then we can derive the lower bound on the achievable sum rate of \ref{eq:rate_ms} in the following proposition.
\begin{proposition}[\textit{Lower bound of the achievable sum rate for MU-SIMO}]
\label{prop:mu-simo}
    Using ZF detection with finite-precision arithmetic in Rayleigh fading, and provided that $M\geq K+1$, the achievable sum rate for MU-SIMO can be lower bounded by
    \begin{equation}
    \label{eq:lower_bound_musimo}
    \small
        \breve{R}_{\mathrm{MS}} = K\log _2\left( 1+\frac{\rho\left( M-K \right)}{1+(c^{\mathrm{u}})^{2}\left( \rho\left( M-K \right) +1 \right) \varUpsilon(M,K)} \right),
    \end{equation}
    where $\varUpsilon(M,K)=\mathbb{E} \left\{ \left( \kappa _2\left( \mathbf{H}^H\mathbf{H} \right) \right) ^2 \right\}$.
\end{proposition}
\begin{IEEEproof}
    The proof is available in Appendix \ref{app:mu-miso}.
\end{IEEEproof}

Note that $\varUpsilon(M, K)$ is the second-order moment of the condition number of the Wishart matrix $\mathbf{H}^H\mathbf{H}$ and exists according to \cite[Theorem 5]{5474635}. Furthermore, the exact expression for $\varUpsilon(M,K)$ can be derived through the distribution of condition numbers for Wishart matrices \cite{5474635,9580677,0609045}. To illustrate, we present a case study with $K=2$ for elaboration.
\begin{case}[$K=2$]
\label{case:K}
    We now focus on the case of dual Wishart matrices in the massive MIMO system, i.e., $K=2$ and $M\gg K$. Then the probability density function (pdf) of $c = \kappa _2\left( \mathbf{H}^H\mathbf{H} \right)$ is given by \cite{9580677}
    \begin{equation}
    \begin{array}{cc}
       f(c) =K\frac{(c-1)^2c^{M-2}}{(c+1)^{2M}},  &          1\leq c \leq \infty,
    \end{array}
    \end{equation}
    where $K = \frac{\Gamma(2M)}{\Gamma(M)\Gamma(M-1)}$. Then we can obtain $\varUpsilon(M,2)$ as follows:
    \begin{equation}\small
    \begin{aligned}
        \varUpsilon(M,2) &= \int_1^{\infty}c^2f(c)\mathrm{d}c \\
        & = K\int_1^{\infty} (c-1)^2c^{M}(c+1)^{-2M}\mathrm{d}c \\
        & \overset{\left(g\right)}{=} \frac{2\Gamma(2M)\Gamma(M-3)}{\Gamma(M)^2\Gamma(M-1)}{_2\mathcal{F}_1\left( M-3,2M;M;-1 \right)},
    \end{aligned}
    \end{equation}
    where $\left(g\right)$ follows \cite[Eq. (9.111)]{gradshteyn2014table}, and $_2\mathcal{F}_1\left( \cdot \right)$ is a hypergeometric function, which can be truncated to a finite number of terms while still yielding good accuracy or obtained by recurrence \cite{0609045}.
\end{case}

\textit{Proposition} \ref{prop:mu-simo} reveals the significance of the channel matrix condition for the MU-SIMO system, in addition to considering the impact of low-precision arithmetic and the number of antennas $M$. 

Notably, in the asymptotic case when employing full-precision arithmetic, i.e., $u\rightarrow 0$, \eqref{eq:lower_bound_musimo} converges to the specific case of a Rayleigh fading channel, as presented in \cite[Eq. (20)]{6457363}. 

Moreover, similar to \textit{Corollary} \ref{coro:impact_M_simo}, in the asymptotic case where $M$ goes without bound, i.e., $M\rightarrow\infty$, \eqref{eq:lower_bound_musimo} tends to $0$, which means that as the number of BS antennas increases, rounding errors will accumulate, resulting in a persistent degradation of communication performance in the MU-SIMO system.

% \begin{corollary}[\textit{Impact of $M$ for MU-SIMO}]
%     \label{coro:M_musimo}
%     For a fixed $\rho$, $K$ and fixed precision arithmetic, when $M$ goes without bound, \eqref{eq:lower_bound_musimo} tends to
%     \begin{equation}
%     \label{eq:impact_miso_m}
%          \lim_{M \rightarrow \infty} R_{\mathrm{MS}}^{l} = 0.
%     \end{equation}
% \end{corollary}
% \begin{IEEEproof}
%    For $M\rightarrow \infty$, $\varUpsilon(M,K)\rightarrow 1$. Note that the numerator and denominator in \eqref{eq:lower_bound_musimo} are the order of $O(M)$ and $O(M^2)$, respectively. Therefore, we have \eqref{eq:impact_miso_m}.
% \end{IEEEproof} 
% Similar to \textit{Corollary} \ref{coro:impact_M_simo}, \textit{Corollary} \ref{coro:M_musimo} shows that as the number of BS antennas increases, rounding errors will accumulate, resulting in a persistent degradation of communication performance in the MU-SIMO system.

\subsection{MU-MISO Using ZF Precoding with Finite-Precision Arithmetic}

\begin{algorithm}[t]
    \caption{NE method-based ZF precoding for MU-MISO systems with finite-precision arithmetic}%标题
    \label{alg:ne_miso}
      \LinesNumbered
      \KwIn {Channel matrix $\mathbf{H}$, transmitted signals $\mathbf{x}^{\mathrm{d}}$.}
      \KwOut {${\bf s}^{(l)}$.}
        Compute the matrix-matrix products 
        \begin{equation}\setlength\abovedisplayskip{1.5pt}
            \setlength\belowdisplayskip{1.5pt}
            {\bf C}^{(l)}=\boldsymbol{fl}\left(\mathbf{H}^H\mathbf{H}\right). \notag
        \end{equation}

        Compute the Cholesky factorization 
        \begin{equation}\setlength\abovedisplayskip{1.5pt}
            \setlength\belowdisplayskip{1.5pt}
            {\bf C}^{(l)}=({\bf R}^{(l)})^H{\bf R}^{(l)}. \notag
        \end{equation}

        Solve the triangular equations
        \begin{equation}\setlength\abovedisplayskip{1.5pt}
            \setlength\belowdisplayskip{1.5pt}
            \begin{array}{cc}
              ({\bf R}^{(l)})^H{\bf q}^{(l)}=\mathbf{x}^{\mathrm{d}},   &  {\bf R}^{(l)}{\bf e}^{(l)}={\bf q}^{(l)}. \notag
            \end{array}
        \end{equation}

        Compute the matrix-vector products 
        \begin{equation}\setlength\abovedisplayskip{1.5pt}
            \setlength\belowdisplayskip{1.5pt}
            \begin{array}{cc}
              {\bf s}^{(l)}=\boldsymbol{fl}\left(\mathbf{H}{\bf e}^{(l)}\right). \notag
            \end{array}
        \end{equation}
\end{algorithm}

We employ ZF precoding in the downlink with the precoding matrix denoted as ${\bf P}={\bf H}(\mathbf{H}^H\mathbf{H})^{-1}$. The transmitted vector after precoding, represented by $\bf s$ is given by
\begin{equation}
    {\bf s} = \sqrt{\beta}{\bf H}\left(\mathbf{H}^H\mathbf{H}\right)^{-1}\mathbf{x}^{\mathrm{d}} \overset{\Delta}{=}\sqrt{\beta}{\bf H}{\bf e},
\end{equation}
where ${\bf e} = \left(\mathbf{H}^H\mathbf{H}\right)^{-1}\mathbf{x}^{\mathrm{d}}\in \mathbb{C}^{K\times 1}$. Follow a similar procedure akin to ZF detection in the MU-SIMO system. First, we solve $\left(\mathbf{H}^H\mathbf{H}\right){\bf e} = \mathbf{x}^{\mathrm{d}}$ to obtain the temporary vector ${\bf e}$. Then, the desired transmitted signals are given by ${\bf s} ={\bf H}{\bf e}$. The entire process is summarized in \textit{Algorithm} \ref{alg:ne_miso}. Moreover, we can derive the rounding error bound for MU-MISO in the following lemma.

\begin{lemma}[\textit{Error bound for MU-MISO}]
\label{lem:MU-MISO}
    When ZF precoding is carried out in the finite-precision arithmetic based on \textit{Algorithm} \ref{alg:ne_miso},  and provided that $c_1^{\mathrm{d}}\kappa _2\left( \mathbf{H}^H\mathbf{H} \right)<1$while ensuring the backward stability of \textit{Algorithm} \ref{alg:ne_miso}, the transmit vector after ZF precoding at the BS can be expressed as 
    \begin{equation}
    \begin{aligned}
                {\bf s} ^{(l)} & = {\bf s} + \Delta {\bf S} \\
                & =  \sqrt{\beta}{\bf H}\left(\mathbf{H}^H\mathbf{H}\right)^{-1}\mathbf{x}^{\mathrm{d}} + \Delta {\bf S},
    \end{aligned}
    \end{equation}
    and the received signal at the $k$th user is given by 
    \begin{equation}
        y_k^{(l)} = \sqrt{\rho\beta} x_k^\mathrm{d} + \sqrt{\rho}{\bf h}_k^{H}\Delta {\bf S} + n_k,
    \end{equation}
    where $\Delta {\bf S} \in \mathbb{C}^{M\times 1}$ satisfies
    \begin{equation}
       \left\|\Delta {\bf S}\right\| _2  \leq  c^{\mathrm{d}} \left\| \mathbf{s} \right\| _2,
        % \left\| \mathbf{H} \right\| _2\left\| \left(\mathbf{H}^H\mathbf{H}\right)^{-1}\mathbf{x}^{\mathrm{d}} \right\| _2,
        % \left\| \mathbf{H} \right\| _2\left\| \left(\mathbf{H}^H\mathbf{H}\right)^{-1}\mathbf{x}^{\mathrm{d}} \right\| _2,
    \end{equation}
    with
    \begin{equation}\small
     \begin{aligned}
                c^{\mathrm{d}} &= c_1^{\mathrm{d}}\kappa _2\left( \mathbf{H}^H\mathbf{H} \right)+ \sqrt{2K}\gamma _{2K}\left( 1 + c_1^{\mathrm{d}}\kappa _2\left( \mathbf{H}^H\mathbf{H} \right)\right), \\
       c_1^{\mathrm{d}} &= c_1^{\mathrm{u}}  = 2K\left( \gamma _{2M}+\gamma _{6K+1}/\left( 1-2K\gamma _{2K+1} \right)\right).
       % \sqrt{2K}\gamma _{2K} + \left( \sqrt{2K}\gamma _{2K} +1 \right)c_1^{\mathrm{d}}\kappa _2\left( \mathbf{H}^H\mathbf{H} \right)
    \end{aligned}
    \end{equation}
\end{lemma}
\begin{IEEEproof}
    The proof is similar to that of \textit{Lemma} \ref{lem:MU-SIMO}, which is omitted for conciseness.
\end{IEEEproof}
Similar to \textit{Lemma} \ref{lem:MU-SIMO}, \textit{Lemma} \ref{lem:MU-MISO} also shows the impact of the number of antennas $M$ and users $K$, and the channel condition on the rounding error. Notably, in contrast to the single-user scenario, the bound of rounding errors in the multi-user scenario is similar. This similarity arises from the fact that both cases undergo \textit{identical matrix computations}, differing only in their respective \textit{orders}. Simulation in Sec. \ref{sec:sim} will confirm our analysis.

Following \textit{Lemma} \ref{lem:MU-MISO}, the ergodic achievable sum rate of the MU-MISO system is given by 
\begin{equation}
\label{eq:er:mu_miso}
     R_{\mathrm{MM}} = \sum_{k=1}^{K}\mathbb{E} \left\{ \log _2\left( 1+\frac{\rho\beta}{ \rho \left| \mathbf{h}_k^H\Delta{\bf S} \right|^2+1} \right) \right\}.
\end{equation}
Then the lower bound of \eqref{eq:er:mu_miso} is presented in the proposition below.
\begin{proposition}[\textit{Lower bound of the achievable sum rate for MU-MISO}]
\label{prop:mu-miso}
     Using ZF precoding with finite-precision arithmetic in Rayleigh fading, and provided that $M\geq K+1$, the achievable sum rate for MU-MISO can be lower bounded by
     \begin{equation}
     \label{eq:lower_b_mu_miso}
         \breve{R}_{\mathrm{MM}} = K\log _2\left( 1+\frac{\rho\left( M-K \right)}{1+\mathbb{E} \left\{(c^{\mathrm{d}})^{2}\right\} \rho MK } \right).
     \end{equation}
\end{proposition}
\begin{IEEEproof}
    Using Cauchy-Schwartz inequity, \eqref{eq:er:mu_miso} can be expressed as
    \begin{align}
        R_{\mathrm{MM}} &\geq \sum_{k=1}^{K}\mathbb{E}\left\{ \log _2\left( 1+\frac{\rho\beta}{ \rho \left\| \mathbf{h}_k \right\|_2^2\left\| \Delta{\bf S} \right\|_2^2+1} \right) \right\}\\
        &\geq \sum_{k=1}^{K}\log _2\left( 1+\frac{\rho \beta }{ \rho \mathbb{E} \left\{\left\| \mathbf{h}_k \right\|_2^2\left\| \Delta{\bf S} \right\|_2^2 \right \}+1} \right)\\
        &\geq K\log _2\left( 1+\frac{\rho(M-K)}{ \mathbb{E} \left\{(c^{\mathrm{d}})^{2}\right\}\rho MK +1} \right).
    \end{align}
    where $\beta = K/{\mathbb{E}\{\mathrm{tr}({\bf P}{\bf P}^H)\}} =M-K$, $\mathbb{E}\left\{\left\| \mathbf{h}_k \right\|_2^2\right\}=M$, and $\mathbb{E}\left\{\left\| \mathbf{s} \right\|_2^2\right\}=\beta \mathbb{E} \left\{ \mathrm{tr}\left( \left( \mathbf{H}^H\mathbf{H} \right) ^{-1} \right) \right\}=\beta K/(M-K)$.
\end{IEEEproof}
Note that $\mathbb{E} \left\{(c^{\mathrm{d}})^{2}\right\}$ is involved in both the first-order and second-order moments of the condition number of $\mathbf{H}^H\mathbf{H}$. The computation of these moments can be performed using the same methodology as illustrated in \textit{Case} \ref{case:K}, and additional details will not be expounded upon.

\textit{Proposition} \ref{prop:mu-miso} indicates the importance of the condition of the channel matrix for the MU-MISO system, in addition to considering the influence of low-precision arithmetic and the number of antennas $M$. Especially, in the presence of an \textit{ill-conditioned} channel, the communication performance experiences significant degradation.

In the asymptotic case of employing full-precision arithmetic, i.e., as $u \rightarrow 0$, \eqref{eq:lower_b_mu_miso} converges to the specific case of a Rayleigh fading channel, as elucidated in \cite[Eq. (18)]{tan2018multiuser}.

Furthermore, similar to the MU-SIMO system, in the asymptotic case where $M$ goes without bound, i.e., $M\rightarrow\infty$, \eqref{eq:lower_b_mu_miso} tends to $0$. 

Different from the single-user scenario, a direct inspection of \textit{Algorithm} \ref{alg:ne_simo} and \ref{alg:ne_miso} reveals that there is a clear symmetry between ZF detection in the MU-SIMO system and ZF precoding in the MU-MISO system. In the former, the received signal ${\bf z}$ undergoes multiplication by ${\bf H}^H$, resulting in the vector ${\bf c} = {\bf H}^H{\bf z}$, which is then processed using the NE method to estimate the users' signals $\bf r$. In the latter, the NE method is first applied to the users' signals ${\bf x}^{\mathrm{d}}$ in the MU-MISO system, and then the output of the NE method is multiplied by $\bf H$ to obtain the transmitted signals. This highly resembles the well-known uplink-downlink duality in the underlying vector Gaussian broadcast/multiple-access channel \cite{1237143}.

\section{Mixed-precision Arithmetic Transceiver Architecture Design}
\label{sec:com}
As analyzed in Sec. \ref{sec:su} and \ref{sec:mu}, despite its great superiority in speed and energy cost, finite-precision arithmetic, particularly low-precision arithmetic denoted as $u_l$, has to tolerate a large rate loss, especially when the number of BS antennas is substantial. To compensate for the performance gap, we first introduce a transceiver design for massive MIMO systems based on mixed-precision arithmetic where low-precision arithmetic partially, but not completely, replaces high-precision arithmetic denoted as $u_h$. For simplification, we assume that $u_h = u_l^2$, i.e., doubled precision. Then a comprehensive analysis of rounding errors and computational costs is presented to show its superiority. 

\begin{figure*}[t]
    \centering
    \includegraphics[width=1\textwidth]{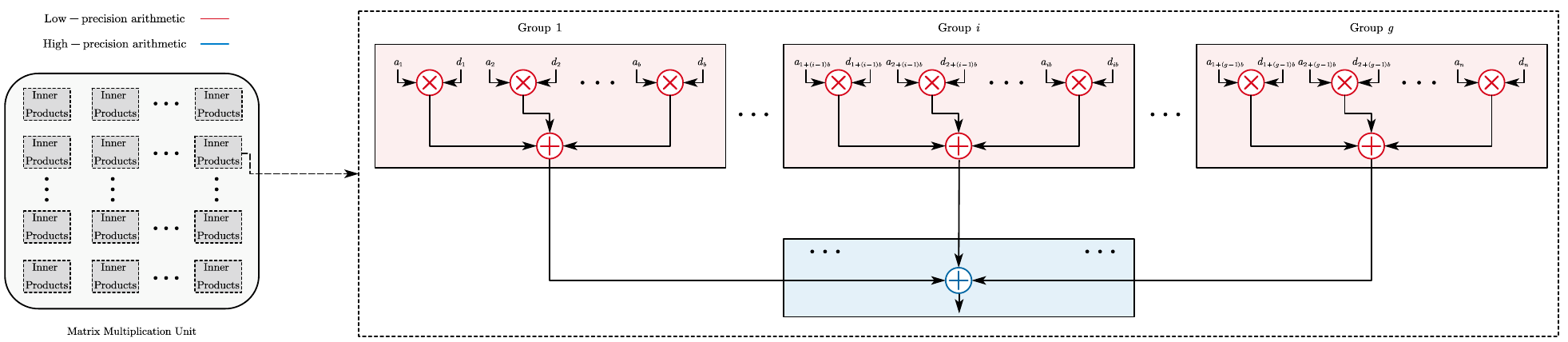}
   \caption{The illustration of mixed-precision arithmetic architecture. This architecture implements the matrix-matrix products based on mixed-precision arithmetic. Each element is computed by the mixed-precision arithmetic-based inner products method. More specifically, we use real-valued inner products, i.e., ${\bf a}^T{\bf d}$, as an example, where ${\bf a,b}\in \mathbb{R}^{n\times 1}$ and $g=\lceil {n}/{b} \rceil$.}
  \label{fig:mp_arith}
\end{figure*}

\subsection{Mixed-Precision Arithmetic Architecture}
Our motivation is that most classic linear algebra computations involve \textit{inner products}, such as matrix-vector or matrix-matrix products, matrix factorization, and solving linear systems. These \textit{inner products} are made up with \textit{scalar multiplication} and \textit{summation}. Given the inherent challenges associated with implementing mixed-precision arithmetic on \textit{scalar multiplication}, our focus lies in its application within the \textit{summation} process.

The most common algorithm to compute $\mathsf{c}=\sum_{i=1}^n \mathsf{w}_i\in \mathbb{R}$ is recursive summation, i.e., start with $\mathsf{c}=\mathsf{w}_1$ and compute $\mathsf{c}= \mathsf{c} + \mathsf{w}_i,\,\,i=2,\cdots,n$. From \cite[Sec. 4.2]{higham2002accuracy}, the computed result in finite-precision arithmetic satisfies
\begin{equation}
    \mathsf{c}^{(l)} = \sum_{i=1}^n \mathsf{w}_i(1+\zeta_i),
\end{equation}
where 
\begin{equation}\label{eq:error_sum}
    |\zeta_i| \leq \gamma_{n-1} = \lambda\sqrt{n-1}u_l + O\left(u_l^2\right).
\end{equation}

To prevent the rounding error from growing with $n$, our objective is to make the first term of \eqref{eq:error_sum} which is the first order of $u_l$, free of $n$. The reason why we neglect the second or higher-order term of \eqref{eq:error_sum} is that $u_l^2$ or higher power of $u_l$ are diminutive constants that can counteract the effect of $n$. 

Note that the blocked summation algorithm \cite{higham2002accuracy}, which is widely used in numerical linear algebra, calculates the sum by grouping $\mathsf{c} = \sum_{i=1}^n \mathsf{w}_i$ into blocks of size $b$. The computation between partial sums remains independent until consolidation into the final result. By doing so, the rounding error bound of partial sums is $\lambda\sqrt{b-1}u_l + O\left(u_l^2 \right)$, because rounding errors incurred in different blocks do not accumulate. Then, the last $n/b-1$ additions, i.e., the combination of partial sums, contribute to the error term $\lambda\sqrt{n/b-1}u_l + O\left(u_l^2\right)$. Therefore, we can compute \textit{partial sums} with \textit{low-precision} arithmetic and then \textit{combine} them to obtain $\mathsf{c}$ using \textit{high-precision} arithmetic, ensuring the first-order term is independent of $n$. In other words, the mixed-precision arithmetic inner products design is done, which is summarized in Fig. \ref{fig:mp_arith}. Furthermore, we can replace all the computing of inner products involved in the process of transceiver realization with the proposed mixed-precision arithmetic architecture.

\subsection{Performance Analysis}
\label{sec:rea}
\subsubsection{Rounding Error Analysis} We give the rounding error analysis of inner products in the following theorem as an example.  To our knowledge, the derived error bound for complex-valued mixed-precision arithmetic is new.
\begin{theorem}[\textit{Complex-valued inner products with mixed-precision arithmetic}]
\label{the:mixed}
    Let ${ \mathsf{c}}=\mathbf{a}^{H}\mathbf{d}$, where $\mathbf{a,d}\in \mathbb{C}^{n\times 1}$, be evaluated in the mixed-precision arithmetic with block size $b$. Under the \textit{Assumption} \ref{ass:rv_error}, the computed $\mathsf{c}^{(l)}$ satisfies
    \begin{equation}
    \label{eq:bound_inner_mix}
        \left\| \mathsf{c}^{(l)} -\mathsf{c} \right\|_2 \leq \sqrt{2} \xi_{b,n} \left\| \mathbf{a} \right\|_2\left\| \mathbf{d} \right\|_2,
    \end{equation}
    where
    {\begin{align}
                \xi_{b,n} &= u_l + \gamma_{b-1}^l+\gamma_{2n/b-1}^h+O\left(u_l^2\right) \notag \\
                 & = u_l + \lambda\sqrt{b-1}u_l + O\left(u_l^2\right) + \lambda\sqrt{2n/b-1}u_h \notag \\
                 &+O\left(u_h^2\right)+O\left(u_l^2\right)\notag \\
                 & = \underset{\mathrm{the \, first \, order\, term}}{\underbrace{\left(\lambda\sqrt{b-1}+1 \right)u_l}} +\underset{\mathrm{the \, second \, order\, term}}{\underbrace{\lambda\sqrt{2n/b-1}u_l^2}} + O\left(u_l^2\right). 
    \end{align}}
        % \xi_{b,n} = \sqrt{2}\gamma_2 + \gamma_{b-1}^l+\gamma_{n/b-1}^h+O(u_l^2).
    % where $\gamma_{b-1}^l=\lambda\sqrt{b-1}u_l+O(u_l^2)$ and $\gamma_{2n/b-1}^h = \lambda\sqrt{2n/b-1}u_h+O(u_h^2) = \lambda\sqrt{2n/b-1}u_l^2+O(u_l^3)$.
\end{theorem}
\begin{IEEEproof}
    The proof is available in Appendix \ref{app:mp}. 
\end{IEEEproof}
\textit{Theorem} \ref{the:mixed} reveals that we can obtain an error bound with a first-order term independent of $n$ by using mixed-precision arithmetic. This is because rounding errors incurred in different blocks do not accumulate and high-precision arithmetic offsets the effect of $n$ on the first-order term. Moreover, in comparison to \textit{Theorem} \ref{the:inner products}, it becomes evident that mixed-precision arithmetic significantly enhances computational performance.% and much tighter than the bound in \cite[Theorem 4.1]{blanchard2020class}. Specifically, the derived error bound grows linearly with $\sqrt{b}$, while the previous bound grows linearly with ${b}$.

Furthermore, it is intuitive to note that the proposed mixed-precision architecture yields superior communication performance for massive MIMO systems compared with pure low-precision arithmetic. For $M = 1000$, $\lambda=1$, $b=32$ and using $\mathtt{fp16}$ arithmetic, i.e., $u_l=4.88\times 10^{-4}$, the relative error $\frac{\left\| \mathsf{c}^{(l)} -\mathsf{c} \right\|_2}{\left\| \mathbf{a} \right\|_2\left\| \mathbf{d} \right\|_2}$ is smaller than $4.5\times 10^{-3}$.
% offset the impact of a larger number of antennas due to the first order of 

\subsubsection{Computational Cost Analysis} We now analyze the computational cost $\mathcal{C}_m$ in the mixed-precision arithmetic architecture. Let $\mathcal{C}_S^l$, $\mathcal{C}_S^h$ and $\mathcal{C}_M^l$ be the cost of the real-valued summation for low-precision arithmetic, high-precision arithmetic, and the cost of the real-valued multiplication for low-precision arithmetic, respectively. Then for complex-valued matrix-matrix products $\bf C=AB$, where $\mathbf{A}\in \mathbb{C}^{m\times n}$, $\mathbf{B}\in \mathbb{C}^{n\times p}$, we have
\begin{equation}\label{eq:cost_1}
    \mathcal{C}_m = 4mp\left\{\frac{2n}{b}\mathcal{C}_S^l\left(b\right)+\mathcal{C}_S^h\left(\frac{2n}{b}\right),\,\, \frac{2n}{b}\mathcal{C}_M^l\left(b\right)\right\}.
\end{equation}
For simplicity, we denote the number of summation and multiplication as the computational cost, i.e., 
\begin{equation}\label{eq:cost_2}\small
  \mathcal{C}_S^l\left(b\right)=b-1, \mathcal{C}_S^h\left(\frac{2n}{b}\right)=G\left(\frac{2n}{b}-1\right),  \mathcal{C}_M^l\left(b\right)=Gb,
\end{equation}
where $G$ is a constant factor. For example, we regard single-precision and half-precision as high-precision arithmetic and low-precision arithmetic, respectively. Then $G = 2$. Substituting \eqref{eq:cost_2} into  \eqref{eq:cost_1}, we obtain
\begin{equation}
    \mathcal{C}_m = 4mp\left\{\left(\frac{2n}{b}\right)\left(G-1\right)+2n-G,\,\, 2n\right\}.
\end{equation}

Similarly, the computational costs of pure low-precision arithmetic $\mathcal{C}_l$ and pure high-precision arithmetic $\mathcal{C}_h$ are given by
\begin{align}
    \mathcal{C}_l &= 4mp\left\{2n-1,\,\, 2n\right\}.\\
     \mathcal{C}_h &= 4mp\left\{G\left(2n-1\right),\,\, 2Gn\right\}.
\end{align}
It is shown that $\mathcal{C}_l$ and $\mathcal{C}_h$ are special case of $\mathcal{C}_m$ with $\mathcal{C}_m(G=1)$ and $\mathcal{C}_m(b=1)$, respectively. For $n=1000$, $b=32$, and $G=2$, we need an extra overhead of only $3.08\%$ compared with pure low-precision arithmetic.

Overall, we can conclude that the proposed mixed-precision arithmetic architecture provides a favorable balance between performance and computational cost.

\section{Simulation Results and Discussion}
\label{sec:sim}
In this section, we will provide numerical results to verify our derived results. First, we will give the simulation setups. Then, we will evaluate the derived lower bound to show the effect of the various system parameters. Finally, the performance and accuracy of the mixed-precision arithmetic architecture is assessed.

\begin{figure}[t]
    \centering
    \subfloat[Rate of the SIMO systems versus $M$ with finite-precision arithmetic and $\rho = 10 \mathrm{dB}$.]{\includegraphics[width=0.35\textwidth]{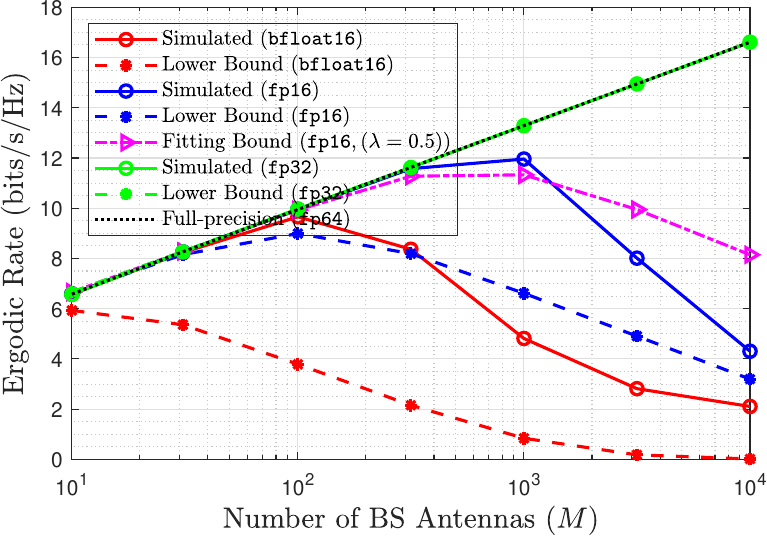}
    \label{fig:simo_m}}
    \vfill
    \subfloat[Rate of the SIMO systems versus $\rho$ with finite-precision arithmetic and $M = 100$.]{\includegraphics[width=0.35\textwidth]{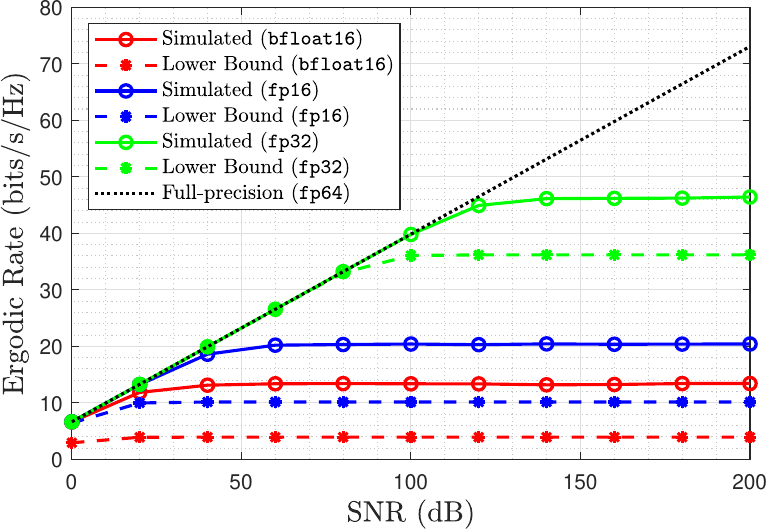}
    \label{fig:simo_rho}}
    \caption{Rate of the SIMO systems with finite-precision arithmetic.}
    \label{fig:simo_simu}
\end{figure}

\subsection{Simulation Setup}
\subsubsection{Simulating Finite-Precision Arithmetic} The authors in \cite{higham2019simulating} provided a MATLAB function, i.e., $\mathtt{chop.m}$, that can be utilized to simulate $\mathtt{fp16}$, $\mathtt{bfloat16}$, and other low-precision arithmetic. Additionally, the implementation of \textit{Algorithm} \ref{alg:ne_simo} and \ref{alg:ne_miso} involves Cholesky factorization and triangular equations. These operations can be realized using the approaches outlined in \cite[Alg. 4.2.2]{gloub1996matrix} and \cite[Alg. 3.1.1\& 3.1.2]{gloub1996matrix}, respectively.

\subsubsection{Simulation Parameters} In the context of finite-precision arithmetic, we define $\mathtt{fp64}$ as full-precision arithmetic, $\mathtt{fp32}$ as high-precision arithmetic, and $\mathtt{fp16}$ and $\mathtt{bfloat16}$ as low-precision arithmetic. For the single-user scenario, we set $\lambda=3$. For the multi-user scenario, we fix $\lambda = 1$, the number of users at $K=4$, and the SNR at $\rho = 10\mathrm{dB}$. Concerning mixed-precision arithmetic, we choose a block size of $b = 32$ and the SNR of $\rho = 10\mathrm{dB}$. 

\subsection{Single-User Scenario}
 \begin{figure}[t]
    \centering
    \subfloat[Rate of MISO systems versus $M$ with finite-precision arithmetic and $\rho = 10 \mathrm{dB}$.]{\includegraphics[width=0.35\textwidth]{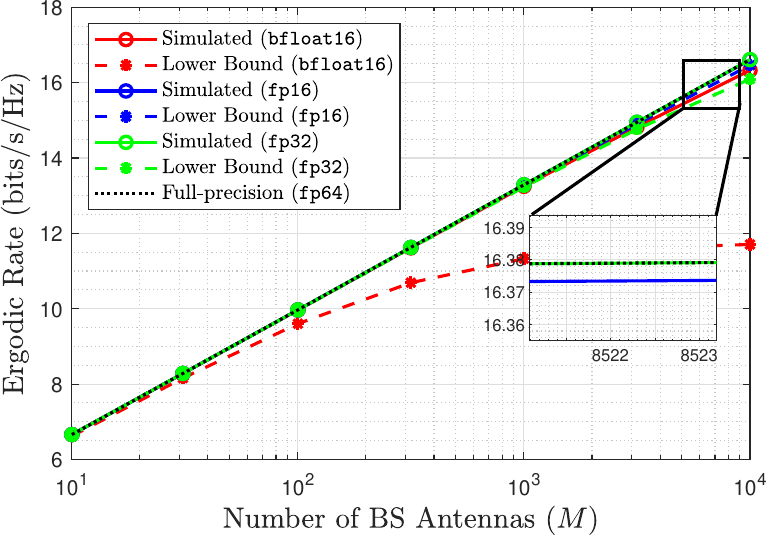}
    \label{fig:miso_m}}
    \vfill
    \subfloat[Rate of MISO systems versus $\rho$  with finite-precision arithmetic and $M = 100,1000,10000$.]{\includegraphics[width=0.35\textwidth]{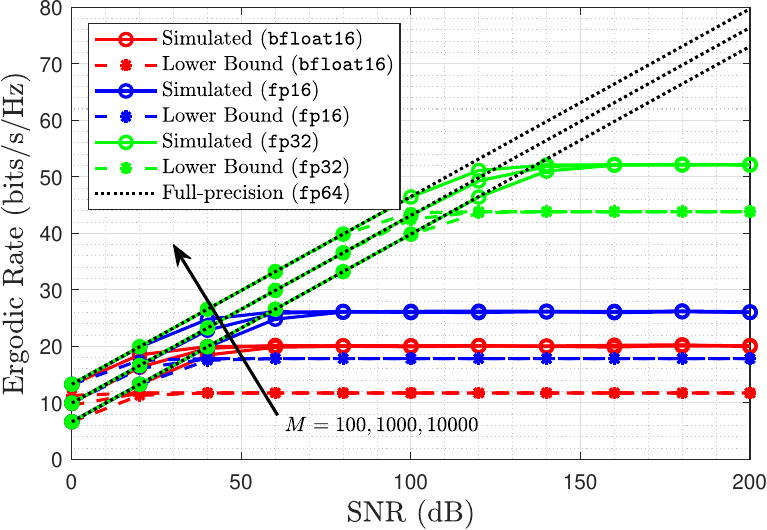}
    \label{fig:miso_rho}}
    \caption{Rate of MISO systems with finite-precision arithmetic.}
    \label{fig:miso_simu}
\end{figure}

\begin{figure}[t]
    \centering
    \subfloat[Rate of MU-SIMO systems versus $M$ with finite-precision arithmetic, $K=4$ and $\rho = 10 \mathrm{dB}$.]{\includegraphics[width=0.35\textwidth]{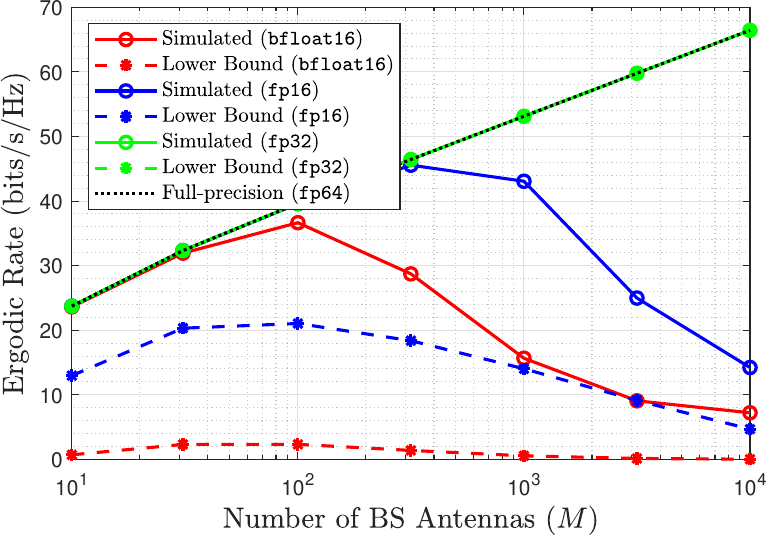}
    \label{fig:musimo_m}}
    \vfill
    \subfloat[Rate of MU-MISO systems versus $M$ with finite-precision arithmetic, $K=4$ and $\rho = 10 \mathrm{dB}$.]{\includegraphics[width=0.35\textwidth]{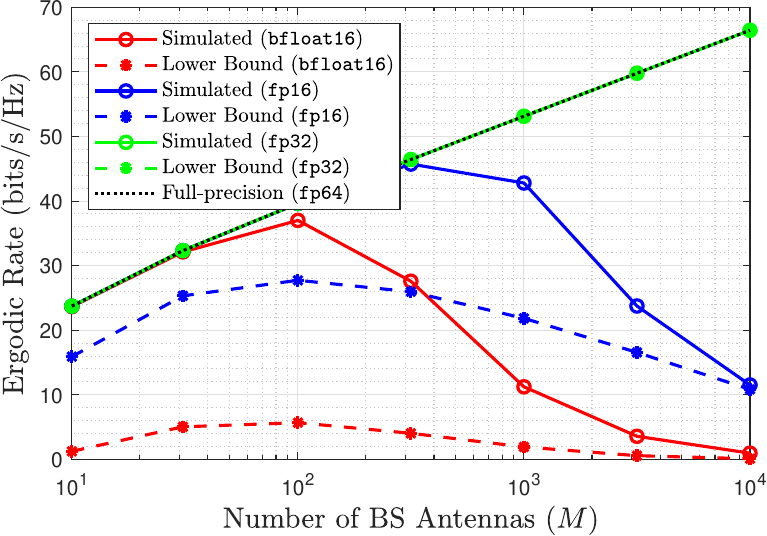}
    \label{fig:mumiso_m}}
    \caption{Rates of MU-SIMO and MU-MISO systems with finite-precision arithmetic.}
    \label{fig:mu_simo_miso_simu}
\end{figure}

First, as shown in Fig. \ref{fig:simo_simu} and \ref{fig:miso_simu}, the ergodic rates of SIMO systems and MISO systems with finite-precision arithmetic are illustrated for different parameters. Note that, comparing Fig. \ref{fig:simo_simu} and \ref{fig:miso_simu}, the anticipated duality between SIMO and MISO systems in the context of finite-precision arithmetic does not hold. We further show the relationship between the ergodic rate and key parameters in the following insights:
\subsubsection{Impact of $u$} It is evident that the ergodic rate of SIMO systems significantly degrades with decreasing arithmetic precision in Fig. \ref{fig:simo_m}. In contrast, the ergodic rate of MISO systems experiences a more tolerable decline as arithmetic precision decreases in Fig. \ref{fig:miso_m}. This discrepancy arises due to their involvement in distinct finite-precision arithmetic, resulting in varying rounding errors, i.e., $\delta_{\mathrm{SIMO}} = \sqrt{2} \gamma_{2M}$ for SIMO and $\delta_{\mathrm{MISO}} = \sqrt{2} \gamma_{2}$ for MISO. Moreover, it can be concluded that for SIMO systems, employing $\mathtt{fp32}$ achieves a performance similar to full-precision arithmetic, while for MISO systems, utilizing $\mathtt{bfloat16}$ approaches the performance of full-precision arithmetic.
\subsubsection{Impact of $M$} Fig. \ref{fig:simo_m} demonstrates that the ergodic rate of SIMO systems exhibits an initially increasing and then decreasing pattern concerning the number of antennas $M$ (refer to \textit{Corollary} \ref{coro:impact_M_simo}). As depicted in Fig. \ref{fig:miso_rho}, for MISO systems, the ergodic rates with different $M$ approach the same value as $\rho$ increases, which validates that the rounding error of MISO systems is independent with $M$ (see \textit{Lemma} \ref{lem:miso}).
\subsubsection{Impact of $\rho$} Fig. \ref{fig:simo_rho} reveals that the ergodic rate of SIMO systems converges to a stable value, validating the accuracy of \textit{Corollary} \ref{coro:impact_rho_simo}. Likewise, in Fig. \ref{fig:miso_rho}, for the case of MISO systems, the ergodic rate also converges to an exact value, which confirms the correctness of \textit{Corollary} \ref{coro:rho_miso}. Note that the values in SIMO and MISO systems are different, which demonstrates that the duality between SIMO and MISO systems is not true in finite-precision arithmetic.

Notably, the derived lower bound is loose but shows the worst-case communication performance. The bound itself is weaker than it might have been because of the necessity of restricting the mass of detail to a reasonable level and because of the limitations imposed by expressing the errors in terms of matrix norms \cite{higham2022mixed,wilkinson1971modern}. Rounding error analysis primarily serves the purpose of uncovering potential instabilities in algorithms, aiming to refine them based on the intrinsic properties obtained \cite{wilkinson1971modern}. Furthermore, following the work of \cite{higham2019new}, we can set a small value $\lambda$ to fit the real rounding error but it is not a lower bound. For example, as illustrated in Fig. \ref{fig:simo_m}, the curve of `Fitting bound $\mathtt{fp16},(\lambda = 0.5)$' approaches closely to the curves of `Simulated' in the small and medium $M$ but exceeds it in the large $M$. 

%This is because the fitting bound is probabilistic and may fail due to a small value $\lambda$.
% Notably, the derived lower bound is not perfectly tight but is acceptable. This is because the primary purpose of rounding error analysis is to reveal potential instabilities in algorithms, with the hope of refining the algorithm based on the intrinsic properties obtained \cite{wilkinson1971modern}, and the derived bound is often more conservative than what it could achieve \cite{higham2022mixed,wilkinson1971modern}. Despite its inherent pessimism, we can leverage the lower bound for a theoretical analysis of the impact of finite-precision arithmetic in massive MIMO systems.

\subsection{Multi-User Scenario}
Then, as depicted in Fig. \ref{fig:mu_simo_miso_simu}, the ergodic rates of MU-SIMO and MU-MISO systems under finite-precision arithmetic are presented for various parameters. It is intuitive to observe that the curves of `Simulated' for MU-SIMO and MU-MISO systems exhibit remarkable similarity, which confirms the duality between MU-SIMO and MU-MISO systems. Further insights are provided below:
\subsubsection{Impact of $u$} Similar to the case of SIMO systems, the ergodic rates of MU-SIMO and MU-MISO systems significantly degrade with decreasing arithmetic precision. Moreover, we can conclude that we can utilize $\mathtt{fp32}$ to achieve a performance similar to full-precision arithmetic for MU-SIMO and MU-MISO systems.
\subsubsection{Impact of $M$} The ergodic rates of MU-SIMO and MU-MISO systems are the first increasing and then decreasing function of $M$ which finally tends to $0$. This is because the rounding errors will accumulate as the number of antennas $M$ grows.

\subsection{Mixed-Precision Arithmetic Architecture}

\begin{figure}[t]
    \centering
    \subfloat[Rate of SIMO systems versus $M$ with mixed-precision arithmetic architecture, $\rho = 10 \mathrm{dB}$ and $b=32$.]{\includegraphics[width=0.35\textwidth]{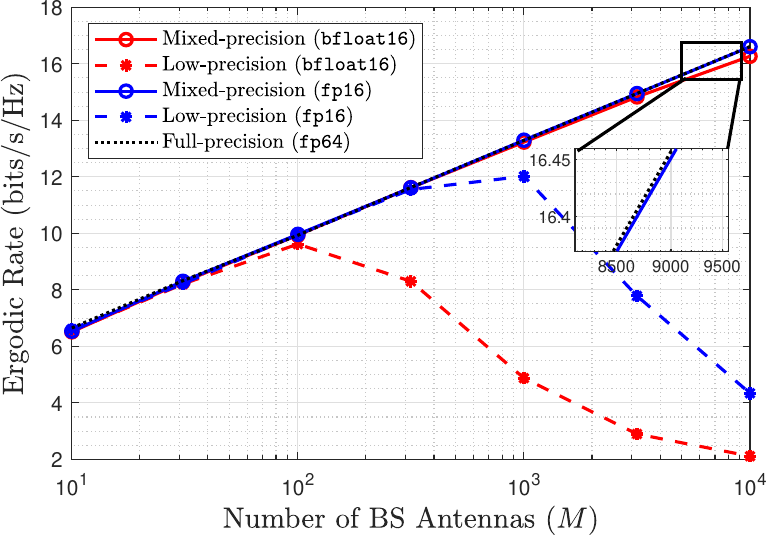}
    \label{fig:mp_simo_m}}
    \vfill
    \subfloat[Rate of MU-SIMO systems versus $M$ with mixed-precision arithmetic architecture, $K=4$, $\rho = 10 \mathrm{dB}$ and $b=32$.]{\includegraphics[width=0.35\textwidth]{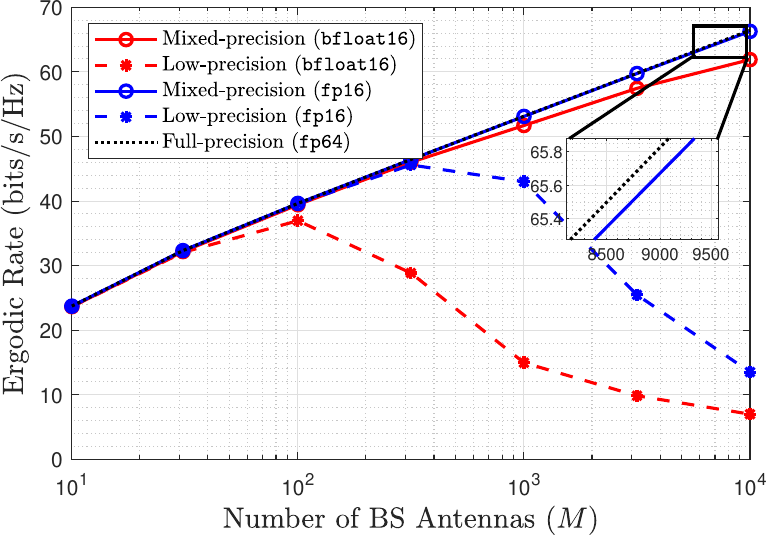}
    \label{fig:mp_mu_simo_m}}
    \caption{Rates of SIMO and MU-SIMO systems with mixed-precision arithmetic architecture.}
    \label{fig:mp_su_musimo_simu}
\end{figure}

 \begin{figure}[t]
     \centering
     \includegraphics[width=0.35\textwidth]{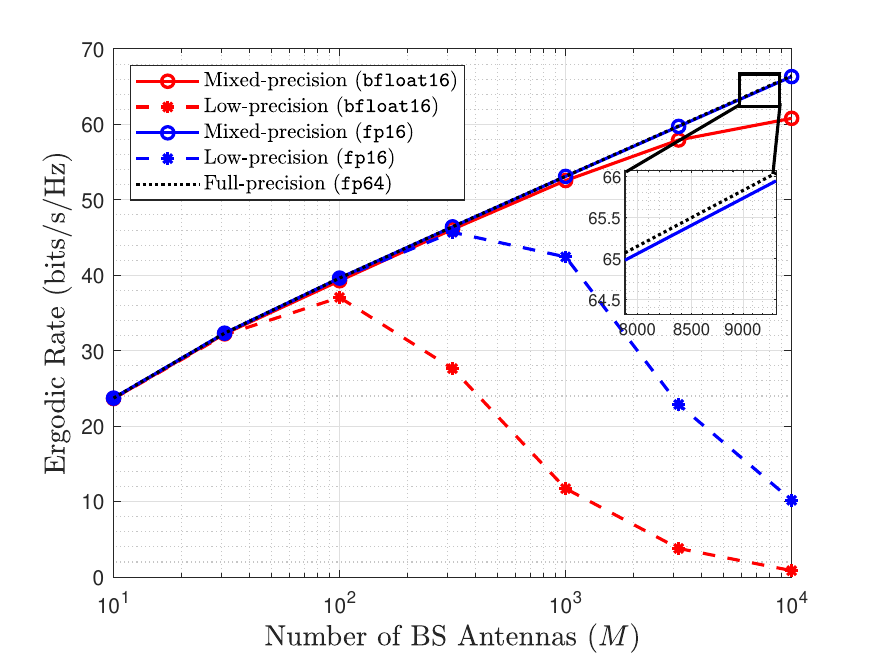}
     \caption{Rate of the MU-MISO systems versus $M$ with mixed-precision arithmetic architecture, $K=4$, $\rho = 10 \mathrm{dB}$ and $b=32$.}
     \label{fig:mu_miso_mixed}
 \end{figure}

Finally, we demonstrate the superiority of the proposed mixed-precision arithmetic architecture under various conditions, including different values of $M$ and $K$, high SNR levels, computational cost, imperfect channel state information (CSI), and other advanced transceivers.
 
\subsubsection{Impact of $M$} We present an illustrative example involving SIMO, MU-SIMO, and MU-MISO systems with ZF detection/precoding to highlight the influence of the proposed mixed-precision arithmetic architecture. Fig. \ref{fig:mp_su_musimo_simu} and Fig. \ref{fig:mu_miso_mixed} demonstrate the rates of SIMO, MU-SIMO, and MU-MISO systems using the proposed mixed-precision arithmetic architecture versus $M$. The results indicate a significant enhancement in communication performance compared with only low-precision arithmetic, approaching levels similar to those achieved with pure full-precision arithmetic. Comparing Fig. \ref{fig:mp_simo_m} and Fig. \ref{fig:mp_mu_simo_m}, we can find that a larger performance gap between mixed-precision and full-precision in MU-SIMO systems since ZF detection in MU-SIMO systems involves more matrix computations than SIMO systems. Despite this, mixed-precision architecture has significantly compensated for the impact of low-precision arithmetic in MU-SIMO systems.

 \begin{figure}[t]
     \centering
     \includegraphics[width=0.35\textwidth]{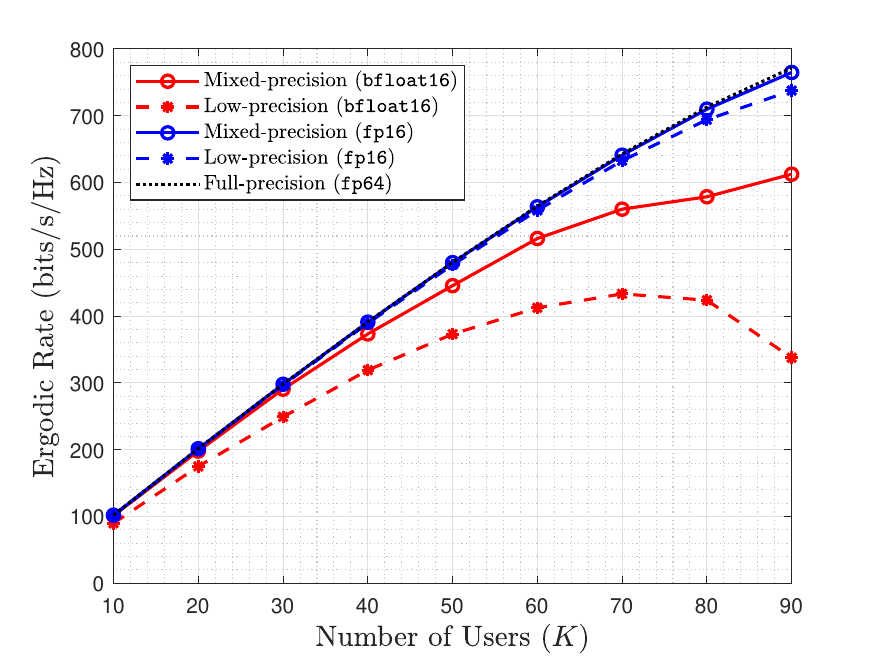}
     \caption{ Rate of MU-SIMO systems versus $K$ with mixed-precision arithmetic architecture, $M=128$, $\rho = 10 \mathrm{dB}$ and $b=8$.}
     \label{fig:mu-simo-k-mp}
 \end{figure}
 
\subsubsection{Impact of $K$} Fig. \ref{fig:mu-simo-k-mp} compares the impact of different finite-precision arithmetic on the communication performance of MU-SIMO systems with varying numbers of users. Lower-precision arithmetic is more sensitive to a larger number of users, leading to a greater performance gap compared to full-precision arithmetic. This sensitivity arises because channels with more users have a higher condition number \cite{9007506}, resulting in increased rounding errors (see \eqref{eq:ne_cv}). Furthermore, the proposed mixed-precision arithmetic architecture mitigates the impact of low-precision arithmetic under different channel conditions, as shown in Fig. \ref{fig:mu-simo-k-mp}.

 \begin{figure}[t]
     \centering
     \includegraphics[width=0.35\textwidth]{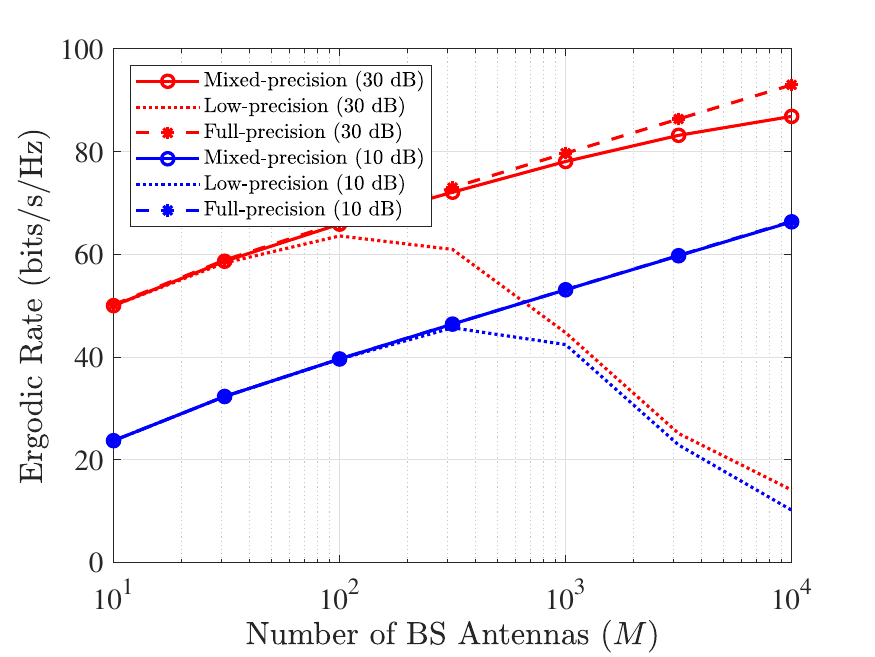}
     \caption{Rate of MU-SIMO systems versus $M$ with mixed-precision arithmetic architecture, $K=4$, $\rho = 10,30 \mathrm{dB}$ and $b=8$.}
     \label{fig:highsnr}
 \end{figure}

\subsubsection{Impact of high SNR $\rho$} As the ZF method demonstrates nearly optimal capacity at high SNR levels, we further present the rate of MU-SIMO systems versus $M$ with mixed-precision arithmetic architecture at $\rho = 30$ dB as an example in Fig. \ref{fig:highsnr}. The results indicate that the proposed mixed-precision arithmetic remains efficient at high SNR levels.

  \begin{figure}[t]
     \centering
     \includegraphics[width=0.35\textwidth]{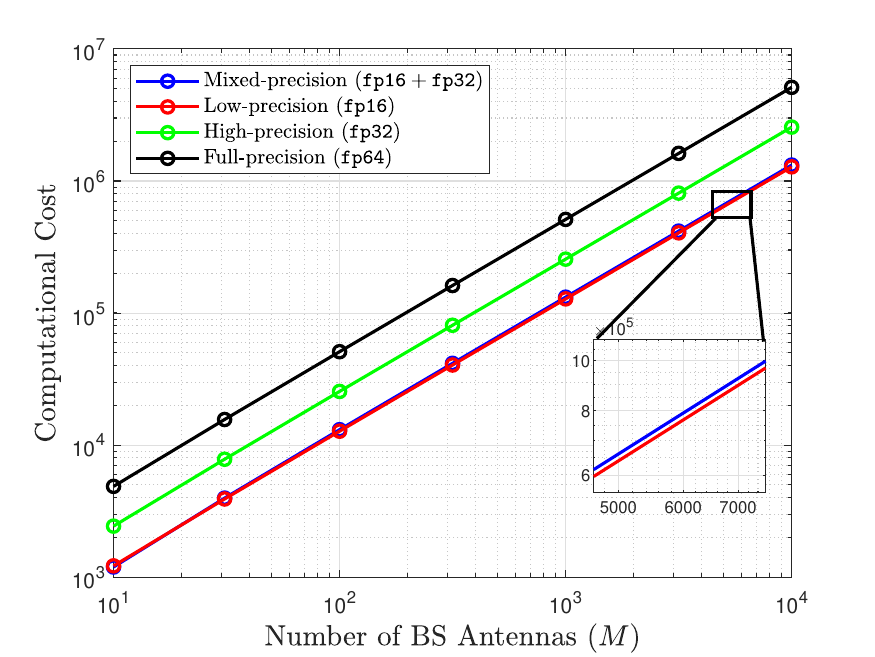}
     \caption{Computational cost of mixed-precision architecture, low-precision arithmetic, and high-precision arithmetic versus $M$ with $b=32$, $G=2$ and $K=4$.}
     \label{fig:c_c}
 \end{figure}

 \begin{figure}[t]
    \centering
    \subfloat[Rate of SIMO systems versus $M$ with mixed-precision arithmetic architecture, $\rho = 10 \mathrm{dB}$ and $b=32$.]{\includegraphics[width=0.35\textwidth]{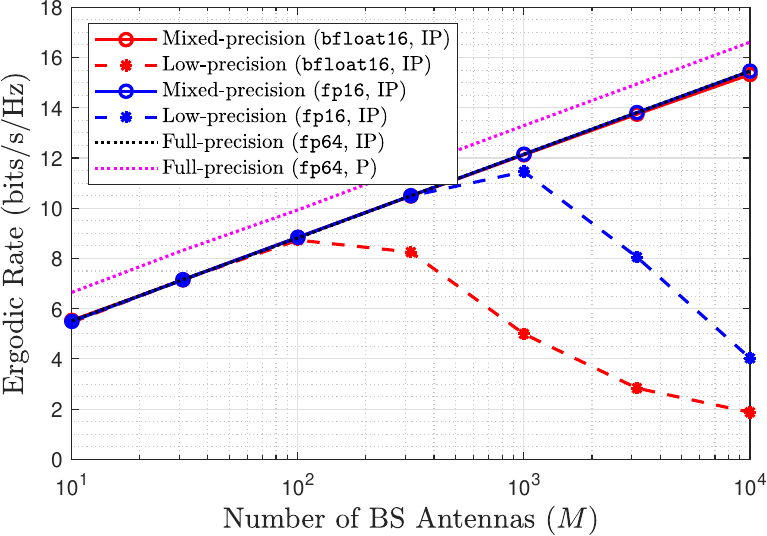}
    \label{fig:mp_simo_m_ip}}
    \vfill
    \subfloat[Rate of MU-SIMO systems versus $M$ with mixed-precision arithmetic architecture, $K=4$, $\rho = 10 \mathrm{dB}$ and $b=32$.]{\includegraphics[width=0.35\textwidth]{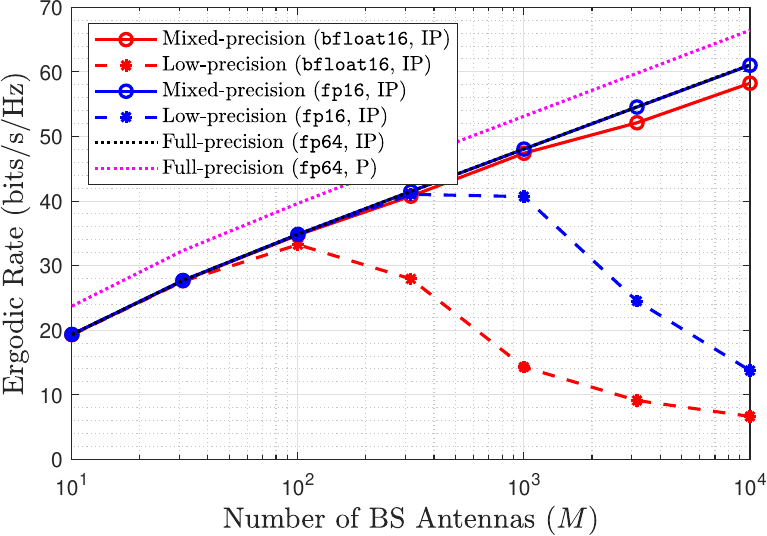}
    \label{fig:mp_mu_simo_m_ip}}
    \caption{Rates of SIMO and MU-SIMO systems with mixed-precision arithmetic architecture and imperfect CSI. 'P' means perfect CSI, and 'IP' means imperfect CSI.}
    \label{fig:mp_su_musimo_simu_ip}
\end{figure}

\subsubsection{Computational Cost} In terms of computational cost, we illustrate with a matrix-matrix product example, i.e., $\mathbf{H}^H\mathbf{H}$, setting $b=32$, $G=2$, and $K=4$. As depicted in Fig. \ref{fig:c_c}, the computational cost of the mixed-precision arithmetic architecture is significantly lower than that of full-precision and high-precision arithmetic, closely approaching that of low-precision arithmetic. Specifically, for $M=1000$, the mixed-precision architecture introduces only a minimal overhead of $3.08\%$ compared with pure low-precision arithmetic. Additionally, it incurs a modest computational overhead of approximately $48.5\%$ and $25.8\%$ when compared with pure high-precision arithmetic and full-precision arithmetic, respectively (see Sec. \ref{sec:rea}).
% The trade-off between performance and computational cost, influenced by the block size $b$, will be explored in future work.

% \begin{figure}[t]
%     \centering
%     \subfloat[Rate of MU-SIMO systems using MMSE detection versus $M$ with mixed-precision arithmetic architecture, $K=4$, $\rho = 10 \mathrm{dB}$ and $b=32$.]{\includegraphics[width=0.3\textwidth]{Figures/MMSE.eps}
%     \label{fig:mu_simo_mmse}}
%     \vfill
%     \subfloat[Rate of MU-MISO systems using WMMSE versus $M$ with mixed-precision arithmetic architecture, $K=16$, $\rho = 10 \mathrm{dB}$, the convergence threshold $\epsilon = 10^{-3}$ and $b=32$.]{\includegraphics[width=0.3\textwidth]{Figures/WMMSE_M.eps}
%     \label{fig:mu-miso-wmmse}}
%     \vfill
%     \subfloat[BER of the MU-SIMO systems using ZF and ZF-SIC versus $M$ with mixed-precision arithmetic, $K = 4$, $b = 32$, SNR = 0 dB, and 16QAM.]{\includegraphics[width=0.3\textwidth]{Figures/BER_ZF_SIC.eps}
%     \label{fig:mu-simo-zf-sic}}
%     \caption{Performance of different transceivers with mixed-precision arithmetic architecture.}
%     \label{fig:advanced_transceivers}
% \end{figure}
\begin{figure*}[t]
    \centering
    \subfloat[Rate of MU-SIMO systems using MMSE detection versus $M$ with mixed-precision arithmetic architecture, $K=4$, $\rho = 10 \mathrm{dB}$ and $b=32$.]{\includegraphics[width=0.3\textwidth]{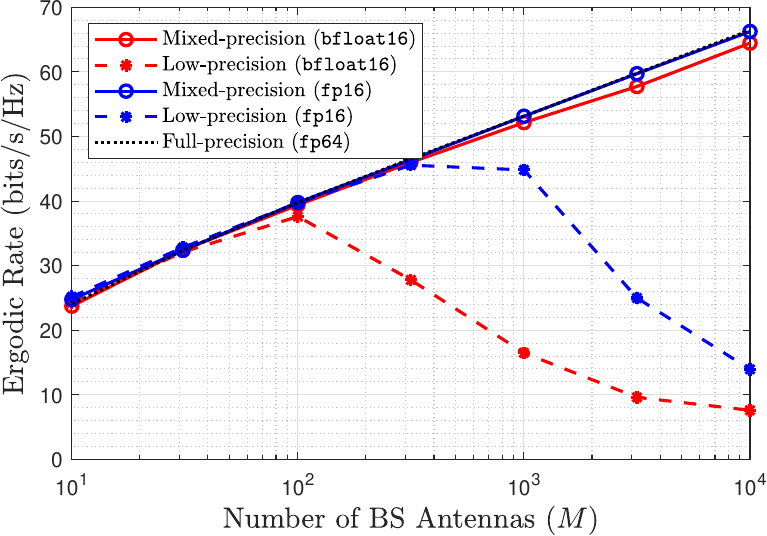}
    \label{fig:mu_simo_mmse}}
    \hfill
    \subfloat[Rate of MU-MISO systems using WMMSE versus $M$ with mixed-precision arithmetic architecture, $K=16$, $\rho = 10 \mathrm{dB}$, the convergence threshold $\epsilon = 10^{-3}$ and $b=32$.]{\includegraphics[width=0.3\textwidth]{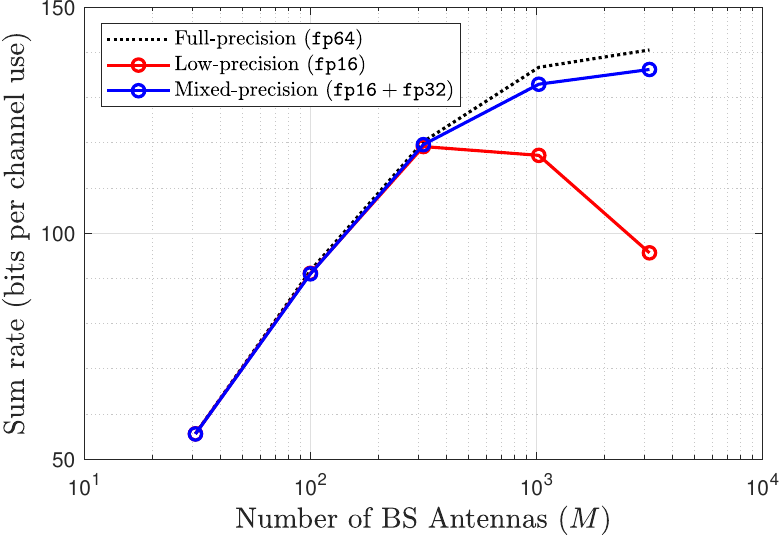}
    \label{fig:mu-miso-wmmse}}
    \hfill
    \subfloat[BER of the MU-SIMO systems using ZF and ZF-SIC versus $M$ with mixed-precision arithmetic, $K = 4$, $b = 32$, SNR = 0 dB, and 16QAM.]{\includegraphics[width=0.3\textwidth]{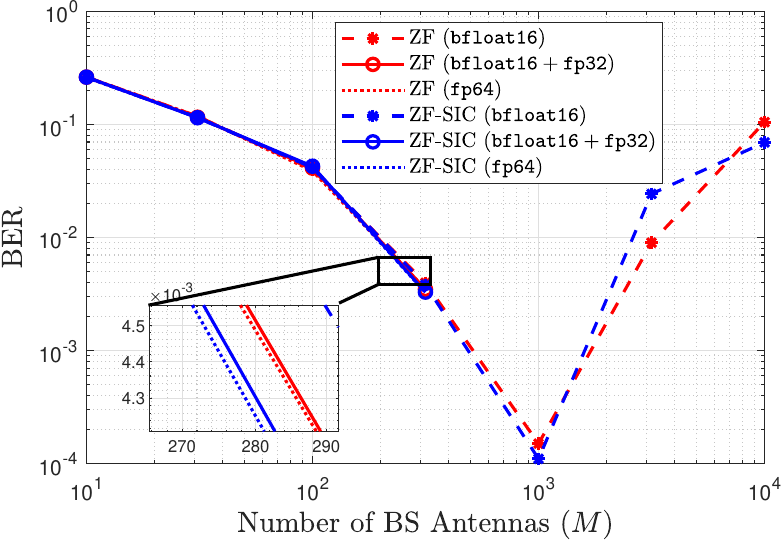}
    \label{fig:mu-simo-zf-sic}}
    \caption{Performance of different transceivers with mixed-precision arithmetic architecture.}
    \label{fig:advanced_transceivers}
\end{figure*}

\subsubsection{Impact of Imperfect CSI} Channel estimation errors are inevitable in practice. Therefore, for the most practical and general case of imperfect CSI, we consider a transmission within the coherence interval $T$ and use $\tau$ symbols for pilots. The CSI for each antenna can be obtained using high-precision arithmetic. The power of pilot symbols is $\rho_p =\tau \rho$, where the MMSE estimate of $\bf H$ is given by $\Hat{\bf H}$. Let $\Delta{\bf H}=\Hat{\bf H}-{\bf H}$ be the channel estimate error matrix, which is independent of $\Hat{\bf H}$. The entries of $\Delta{\bf H}$ are random variables with zero means and variances $\frac{1}{\rho_p+1}$ \cite{6457363}. Following the simulation parameters in \cite{6457363,6816003}, we set the coherence interval $T=196$, and pilot sequences of length $\tau = K$. Other parameters are the same as those in the perfect CSI scenario. 
% Following a similar analysis in perfect CSI and \cite{6457363,6816003}, the ergodic achievable sum rate of the SIMO and MU-SIMO system can be expressed as 
% \begin{equation}
% {\footnotesize
%     R_{\mathrm{SIMO}}^{\mathrm{IP}} = \frac{T-\tau}{T}\mathbb{E}\left\{ \log_2 \left(1+\frac{\rho\left\| \mathbf{\Hat{h}} \right\|_2^4}{\left(1+\frac{\rho}{\rho_p+1}\right)\left\| \mathbf{\Hat{h}} \right\|_2^2+\left\| \Delta r \right\|_2^2} \right)\right\} \notag.}
% \end{equation}
% \begin{equation}
% \footnotesize
%     R_{\mathrm{MS}}^{\mathrm{IP}}  = \frac{T-\tau}{T}\sum_{k=1}^{K}\mathbb{E} \left\{ \log _2\left( 1+\frac{\rho}{\left(\frac{K\rho}{\rho_p+1}+1\right)\left[ \left( \Hat{\bf H}^H\Hat{\bf H} \right) ^{-1} \right] _{kk}+\left\| \Delta{\bf r}_k \right\| _2^2} \right) \right\} \notag.
% \end{equation}
As shown in Fig. \ref{fig:mp_simo_m_ip} and \ref{fig:mp_mu_simo_m_ip}, the ergodic rates of SIMO and MU-SIMO systems under imperfect CSI are presented with different precision levels. It is clear that the proposed mixed-precision arithmetic architecture still performs well between full-precision and low-precision in the imperfect CSI scenario.

\subsubsection{More Advanced Transceivers} More advanced precoding and receiver schemes are presented to demonstrate the superiority of the proposed mixed-precision arithmetic architecture, including linear transceiver methods, such as minimum mean-squared error (MMSE) \cite{4686826}, and non-linear transceiver methods, such as zero-forcing successive interference cancellation (ZF-SIC) \cite{golden1999detection}, and weighted minimum mean-squared error (WMMSE) \cite{5756489}, as depicted in Fig. \ref{fig:advanced_transceivers}. It is clear that the proposed mixed-precision arithmetic architecture consistently performs well in terms of ergodic rate and bit error rate (BER).

\section{Conclusions}
\label{sec:con}
In this paper, we have utilized finite-precision arithmetic to realize low computational complexity massive MIMO transceivers. First, we have derived the rounding error bound and lower bound of the achievable rate for SIMO systems using MRC and MISO systems using MRT with finite-precision arithmetic, respectively. Then, for the multi-user scenario, lower bounds of achievable rates for MU-SIMO and MU-MISO using ZF with finite-precision arithmetic have been derived. Our derivations have unveiled the impact of finite-precision arithmetic on massive MIMO transceivers. Finally, to mitigate the impact of finite-precision arithmetic, particularly low-precision arithmetic, we have proposed a mixed-precision arithmetic architecture, offering a favorable balance between performance and computational cost. Simulation results have illustrated the influence of different system configurations on performance and shown the advantage of the proposed architecture.

\appendices
\section{Proof of \textit{Theorem} \ref{the:inner products}}
\label{app:inner products}
Before beginning the derivation of the proof. We first give the useful lemma regarding norm inequity and identity as follows.
\begin{lemma}[\textit{Norm inequity and identity}{\cite[Lemma 6.6]{higham2002accuracy}}]
\label{lem:inequity}
    Let $\mathbf{A,B}\in \mathbb{R}^{m\times n}$ and $\mathbf{c,d} \in \mathbb{R}^{m}$. 
    
    (a) If $\left| \mathbf{A} \right| \leq \left| \mathbf{B} \right|$, then $\left\| \mathbf{A} \right\|_2 \leq \left\| \mathbf{B} \right\|_2$.  
    
    (b) $\left\| \mathbf{A} \right\|_2\leq \left\| \left| \mathbf{A} \right| \right\|_2\leq \sqrt{\mathrm{rank}(\mathbf{A})}  \left\| \mathbf{A} \right\|_2$.

    (c) If $\left| \mathbf{c} \right| \leq \left| \mathbf{d} \right|$, then $\left\| \mathbf{c} \right\|_2 \leq \left\| \mathbf{d} \right\|_2$.

    (d) $\left\| \mathbf{c} \right\|_2 =  \left\| \left| \mathbf{c} \right| \right\|_2$.
\end{lemma}

Then we transform complex-valued variables into their equivalent real-valued representations by Widely-Linear (WL) notation \cite{8527663}, and have
\begin{align}
   & { s}=\mathbf{a}^{H}\mathbf{b} \in \mathbb{C},\\
    \Longleftrightarrow  & \underset{\tilde{\mathbf{s}}\in \mathbb{R}^{2\times 1}}{\underbrace{\left[ \begin{array}{c}
	\Re \left( {{s}} \right)\\
	\Im \left( {{s}} \right)\\
\end{array} \right] }}=\underset{\tilde{\mathbf{A}}\in \mathbb{R}^{2n\times 2}}{\underbrace{\left[ \begin{matrix}
	\Re \left( {\mathbf{a}} \right)&		-\Im \left( {\mathbf{a}} \right)\\
	\Im \left( {\mathbf{a}} \right)&		\Re \left( {\mathbf{a}} \right)\\
\end{matrix} \right] }}^T\underset{\tilde{\mathbf{b}}\in \mathbb{R}^{2n\times 1}}{\underbrace{\left[ \begin{array}{c}
	\Re \left( {\mathbf{b}} \right)\\
	\Im \left( {\mathbf{b}} \right)\\
\end{array} \right] }}\label{eq:rv_cv}.
\end{align}
Additionally, we use \textit{Lemma} \ref{lem:rv_inner} for \eqref{eq:rv_cv} and have
\begin{equation}
\label{eq:inequity_inner_abs}
    \left| \tilde{\mathbf{s}}^{(l)} - \tilde{\mathbf{s}}\right| \leq \gamma_{2n} \left| \tilde{\mathbf{A}}^T\right|\left| \tilde{\mathbf{b}}\right|,
\end{equation}
where $\tilde{\mathbf{s}}^{(l)}$ is is the result of  finite-precision arithmetic of $\tilde{\mathbf{s}}$. Furthermore, we can obtain the expression of \eqref{eq:inequity_inner_abs} in the norm case through \textit{Lemma} \ref{lem:inequity}, i.e.,
\begin{align}
     \left\| \tilde{\mathbf{s}}^{(l)} - \tilde{\mathbf{s}} \right\|_2 = \left\| \left| \tilde{\mathbf{s}}^{(l)} - \tilde{\mathbf{s}} \right| \right\|_2 &\leq \gamma _{2n}\left\| \left| \tilde{\mathbf{A}}^T \right|\left|  \tilde{\mathbf{b}} \right| \right\|_2 \\
     &\le \gamma _{2n}\left\| \left| \tilde{\mathbf{A}}^T \right| \right\| _2\left\| \left| \tilde{\mathbf{b}}\right| \right\|_2 \\
     &\le \sqrt{2}\gamma _{2n}\left\| \tilde{\mathbf{A}} \right\| _2\left\| \tilde{\mathbf{b}} \right\| _2.\label{eq:rv_inner_complete}
\end{align}
Note that $\left\| \tilde{\mathbf{s}} \right\|_2 = \left\|{{s}} \right\|_2$ and $\| \tilde{\mathbf{A}} \|_2 = \sigma_{\max}(\tilde{\mathbf{A}})= \left\|{\mathbf{a}} \right\|_2$. Thus, \eqref{eq:rv_inner_complete} can be converted into complex-valued representations, i.e., \eqref{eq:bound_inner}, and completes the proof of \textit{Theorem} \ref{the:inner products}.

\section{Proof of \textit{Lemma} \ref{lem:MU-SIMO}}
\label{app:NE_D}
We first give a lemma regarding matrix norm inequity as follows:
\begin{lemma}[\textit{Matrix norm inequity}{\cite[Theorem 2.3.4]{gloub1996matrix}}]
\label{lem:sin_ineq}
    Let full rank ${\bf A,B}\in \mathbb{C}^{n\times n}$. If $\mathrm{rank}(\bf A) = \mathrm{rank}(\bf B)$ and $\eta = \left\| {\bf A}^{-1} \right\|_2\left\| {\bf A}-{\bf B} \right\|_2 < 1$, then
    \begin{equation}
         \left\| {\bf B}^{-1} \right\|_2 \leq \frac{1}{1-\eta} \left\| {\bf A}^{-1} \right\|_2.
    \end{equation}
\end{lemma}

Then using \textit{Theorem} \ref{the:cv_mv_mm}, the rounding error bound for \textit{step} 1 and 2 in \textit{Algorithm} \ref{alg:ne_simo} is given by
{\begin{align}
    &{\bf c}^{(l)}=\mathbf{H}^H\mathbf{z} + \Delta{\bf c}, \,\,\,\,  \left\| \Delta{\bf c} \right\| _2\leq \sqrt{2K} \gamma_{2M} \left\| {\bf H} \right\|_2\left\| {\bf z} \right\|_2. \\
    &{\bf C}^{(l)}=\mathbf{H}^H\mathbf{H} + \Delta{\bf C}_1, \,\,\,\, \left\| \Delta{\bf C}_1 \right\| _2\leq 2K \gamma_{2M} \left\| {\bf H} \right\|_2^2.
\end{align}}
By \cite[Theorem 10.3\&10.4]{higham2002accuracy}, the computed Cholesky factor ${\bf R}^{(l)}$ and solution ${\bf r}^{(l)}$ at \textit{step} 3 and 4 satisfy
{\begin{align}
    &({\bf R}^{(l)})^H{\bf R}^{(l)} = {\bf C}^{(l)}+\Delta{\bf C}_2, \\
    &\left\| \Delta{\bf C}_2 \right\| _2\leq {2K}\gamma_{2K+1}/(1-2K\gamma_{2K+1}) \left\| {\bf H} \right\|_2^2. \\
    &\left({\bf C}^{(l)}+\Delta{\bf C}_3\right){\bf r}^{(l)} = {\bf c}^{(l)},\\
    &\left\| \Delta{\bf C}_3 \right\| _2\leq {2K}\gamma_{6K+1}/(1-2K\gamma_{2K+1}) \left\| {\bf H} \right\|_2^2.
\end{align}}
Overall, we have
{\begin{align}
   &\left({\bf H}^H{\bf H} + \Delta{\bf H}\right)  {\bf r}^{(l)} = {\bf H}^H{\bf z} +\Delta {\bf c}, \\ %\label{eq:abab}
   & \left\| \Delta{\bf H} \right\| _2 = \left\| \Delta{\bf C}_1+\Delta{\bf C}_3 \right\| _2 \notag \\
   &\leq c_1^{\mathrm{u}}\left\| {\bf H} \right\|_2^2 = c_1^{\mathrm{u}}\left\| {\bf H}^H{\bf H} \right\|_2, \\
   & \left\| \Delta{\bf c} \right\| _2 \leq \sqrt{2K} \gamma_{2M} \left\| {\bf H} \right\|_2\left\| {\bf z} \right\|_2,
\end{align}}
where $c_1^{\mathrm{u}}=2K\left( \gamma_{2M} + \gamma_{6K+1}/(1-2K\gamma_{2K+1})\right)$.

Furthermore, let $\mathbf{B}\overset{\Delta}{=}{\bf H}^H{\bf H} + \Delta{\bf H}$ and note that ${\bf H}^H{\bf H} {\bf r} = {\bf H}^H{\bf z}$, the expression for ${\bf r}^{(l)}-{\bf r}$ is
    \begin{align}
        {\bf r}^{(l)}-{\bf r} & = \mathbf{B}^{-1} \left({\bf H}^H{\bf z} +\Delta {\bf c}\right) -{\bf r}\notag \\
        & = \mathbf{B}^{-1}\left({\bf H}^H{\bf H} {\bf r} +\Delta {\bf c}\right) -{\bf r} \notag \\
        & = \mathbf{B}^{-1}\left(\mathbf{B} {\bf r} -\Delta{\bf H}{\bf r} +\Delta {\bf c}\right) -{\bf r} \notag\\
        & = \mathbf{B}^{-1}\left( -\Delta{\bf H}{\bf r} +\Delta {\bf c}\right).
    \end{align}
Applying \textit{Lemma} \ref{lem:sin_ineq} and assuming that the process of the NE method is backward stable, i.e., $\left\| {\bf H} \right\|_2\left\| {\bf z} \right\|_2 \approx \left\| {\bf H}^H{\bf z} \right\|_2$ \cite[Page 387]{higham2002accuracy}, we obtain
\begin{align}
       & \left\| {\bf r}^{(l)}-{\bf r}\right\| _2 \leq \left\| \mathbf{B}^{-1}\right\| _2\left(  \left\|\Delta{\bf H}{\bf r}\right\| _2 +\left\|\Delta {\bf c}\right\| _2\right) \notag\\
        &\leq \frac{\left\| \left({\bf H}^H{\bf H}\right)^{-1}\right\|_2}{1-\left\| \left({\bf H}^H{\bf H}\right)^{-1}\right\|_2\left\| \Delta{\bf H}\right\|_2}\left(  \left\|\Delta{\bf H}\right\| _2\left\|{\bf r}\right\| _2 +\left\|\Delta {\bf c}\right\| _2\right)\notag \\
        &\leq \frac{\left\| \left({\bf H}^H{\bf H}\right)^{-1}\right\|_2}{1-c_1^{\mathrm{u}}\left\| \left({\bf H}^H{\bf H}\right)^{-1}\right\|_2\left\| {\bf H}^H{\bf H} \right\|_2} (c_1^{\mathrm{u}}\left\| {\bf H}^H{\bf H} \right\|_2\left\|{\bf r}\right\| _2 \notag\\
        & + \sqrt{2K} \gamma_{2M} \left\| {\bf H} \right\|_2\left\| {\bf z} \right\|_2)\notag \\
        & = \frac{\left\| \left({\bf H}^H{\bf H}\right)^{-1}\right\|_2}{1-c_1^{\mathrm{u}}\kappa _2\left( \mathbf{H}^H\mathbf{H} \right)}(c_1^{\mathrm{u}}\left\| {\bf H}^H{\bf H} \right\|_2\left\|{\bf r}\right\| _2 \notag \\
        &+ \sqrt{2K} \gamma_{2M} \left\| {\bf H} \right\|_2\left\| {\bf z} \right\|_2)\notag\\
        & \approx \frac{  \kappa _2\left( \mathbf{H}^H\mathbf{H} \right)\left\|{\bf r}\right\| _2 }{1-c_1^{\mathrm{u}}\kappa _2\left( \mathbf{H}^H\mathbf{H} \right)}\left(c_1^{\mathrm{u}} +\sqrt{2K} \gamma_{2M} \frac{\left\| {\bf H}^H{\bf z} \right\|_2}{\left\| {\bf H}^H{\bf H} \right\|_2\left\|{\bf r}\right\| _2}\right)\notag \\
        & \leq \frac{\kappa _2\left( \mathbf{H}^H\mathbf{H} \right)\left\|{\bf r}\right\| _2}{1- c_1^{\mathrm{u}}\kappa _2\left( \mathbf{H}^H\mathbf{H} \right)} \left( c_1^{\mathrm{u}} + \sqrt{2K} \gamma_{2M}   \right) \notag\\
         & = c^{\mathrm{u}}\kappa _2\left( \mathbf{H}^H\mathbf{H} \right)\left\|{\bf r}\right\| _2 +O(u^2) \notag \\
        &\overset{\left(h\right)}{\approx } c^{\mathrm{u}}\kappa _2\left( \mathbf{H}^H\mathbf{H} \right)\left\|{\bf r}\right\| _2 \label{eq:mumiso_proof},
\end{align}
where $c^{\mathrm{u}} = c_1^{\mathrm{u}} + \sqrt{2K} \gamma_{2M}$. In $\left(h\right)$ of \eqref{eq:mumiso_proof}, $c_1^{\mathrm{u}}$ is always a small constant. For $n =1000$, $K = 4$, and using $\mathtt{fp16}$ arithmetic, i.e., $u=4.88\times 10^{-4}$, $c_1^{\mathrm{u}}$ is only $4.6\times 10^{-3}$. And $\kappa _2\left( \mathbf{H}^H\mathbf{H} \right)\rightarrow 1$ for $M\gg K$, so this is not an unfair approximation.

Compared with \cite[Eq. (20.14)]{higham2002accuracy}, we derive the \textit{exact} expression of the constant $c^{\mathrm{u}}$, and the proof of  \textit{Lemma} \ref{lem:MU-SIMO} is done.

\section{Proof of \textit{Proposition} \ref{prop:mu-simo}}
\label{app:mu-miso}
By using Jensen’s inequality and random matrix theory, we obtain the following lower bound on the achievable sum rate:
{\begin{small}
    \begin{align}
          & R_{\mathrm{MS}} \overset{\left( d \right)}{\geq} \sum_{k=1}^{K}\log _2\left( 1+\frac{\rho}{\mathbb{E} \left\{ \left[ \left( \mathbf{H}^H\mathbf{H} \right) ^{-1} \right] _{kk} \right \} +\mathbb{E} \left\{ \left\| \Delta{\bf r}_k \right\| _2^2\right \}} \right) \notag\\
           & \overset{\left( e \right)}{=} K\times\left( \frac{1}{K} \sum_{k=1}^{K}\log _2\left( 1+\frac{\rho}{\frac{1}{M-K} +\mathbb{E} \left\{ \left\| \Delta{\bf r}_k \right\| _2^2\right \}} \right)\right) \notag \\
           & \overset{\left( d \right)}{\geq} K\log _2\left( 1+\frac{\rho}{\frac{1}{M-K} +\frac{1}{K} \sum_{k=1}^{K}\mathbb{E} \left\{ \left\| \Delta{\bf r}_k \right\| _2^2\right \}} \right)\notag \\
           & = K\log _2\left( 1+\frac{\rho}{\frac{1}{M-K} +\frac{1}{K} \mathbb{E} \left\{ \left\| \Delta{\bf r} \right\| _2^2\right \}} \right)\notag \\
           & \overset{\eqref{eq:ne_cv}}{\geq} K\log _2\left( 1+\frac{\rho}{\frac{1}{M-K} +\frac{1}{K} \mathbb{E} \left\{ (c^{\mathrm{u}})^2(\kappa _2\left( \mathbf{H}^H\mathbf{H} \right))^2 \left\| \mathbf{r} \right\| _2^2  \right \}} \right)\notag \\
           & \overset{\left( f\right)}{=} K\log _2\left( 1+\frac{\rho}{\frac{1}{M-K} +\frac{(c^{\mathrm{u}})^2}{K} \varUpsilon(M,K) \left( \rho K+\frac{K}{M-K} \right)   } \right)\notag \\
           & = K\log _2\left( 1+\frac{\rho \left( M-K \right)}{1+(c^{\mathrm{u}})^{2}\left( \rho \left( M-K \right) +1 \right) \varUpsilon(M,K)} \right) ,
    \end{align} 
\end{small}}

where $\left( d \right)$ follows Jensen’s inequality, $\left( e \right)$ follows the identity $\mathbb{E} \left\{ \left[ \left( \mathbf{H}^H\mathbf{H} \right) ^{-1} \right] _{kk} \right \} = \mathbb{E} \left\{ \mathrm{tr}\left[ \left( \mathbf{H}^H\mathbf{H} \right) ^{-1} \right] \right \}/K = 1/(M-K)$ \cite{tulino2004random}, and $\left( f\right)$ follows 
\begin{align}
    \mathbb{E} \left\{ \left\| \mathbf{r} \right\| _{2}^{2} \right\} & =\mathbb{E} \left\{ \left\| \mathbf{A}^H\mathbf{y} \right\|_{2}^{2} \right\} =\mathbb{E} \left\{ \left\| \sqrt{\rho}\mathbf{x}+\mathbf{A}^H\mathbf{n} \right\| _{2}^{2} \right\}\notag \\
    &=\mathbb{E} \left\{ \rho \mathbf{x}^H\mathbf{x} \right\} +\mathbb{E} \left\{ \mathbf{n}^H\mathbf{A}\mathbf{A}^H\mathbf{n} \right\}\notag \\
&=\left( \rho K+\mathbb{E} \left\{ \mathrm{tr}\left( \mathbf{A}\mathbf{A}^H\mathbf{nn}^H \right) \right\} \right) \notag\\
& =\left( \rho K+\mathbb{E} \left\{ \mathrm{tr}\left( \left( \mathbf{H}^H\mathbf{H} \right) ^{-1} \right) \right\} \right)\notag \\
&=\left( \rho K+\frac{K}{M-K} \right)\notag.
\end{align}

\section{Proof of \textit{Theorem} \ref{the:mixed}}
\label{app:mp}
Let $a_i$ and $d_i$ be the $i$th elements of $\mathbf{a}$ and $\mathbf{d}$, respectively. Then we have
\begin{equation}\label{eq:mx}\setlength\abovedisplayskip{3.5pt}
            \setlength\belowdisplayskip{3.5pt}
\begin{aligned}
    a_i^{*} d_i&= \left(\Re\left(a_i \right)\Re\left(d_i \right) +\Im\left(a_i \right)\Im\left(d_i \right) \right)  \\
    &+ \boldsymbol{i}\left(\Re\left(a_i \right)\Im\left(d_i \right) -\Im\left(a_i \right)\Re\left(d_i \right) \right) \\
    & \overset{\Delta}{=} \left(\mathsf{e}_i +\mathsf{e}_{i+1} \right)  
    + \boldsymbol{i}\left(\mathsf{f}_i +\mathsf{f}_{i+1} \right)
\end{aligned}
\end{equation}
where $\mathsf{e},\mathsf{f} \in \mathbb{R}^{2n\times 1}$. In other words, the complex-valued inner products can be converted into two groups of real-valued inner products in parallel. Considering the computing of scalar-scalar products in the low-precision arithmetic, \eqref{eq:mx} can be given by 
{\begin{align}
    \mathsf{e}_i^{(l)}& = \boldsymbol{fl}(\Re\left(a_i \right)\Re\left(d_i \right))  = \mathsf{e}_i\left(1+\delta_{i}\right), \\
    \mathsf{f}_i^{(l)}& = \boldsymbol{fl}(\Re\left(a_i \right)\Im\left(d_i \right))  = \mathsf{f}_i\left(1+\delta_{i}\right).
\end{align}}
where $|\delta_i|\leq u_l$ according to the \textit{Definition} \ref{def:samodel}.

Then each of the partial sums for the real part of $\mathsf{c}_i$ with low-precision arithmetic satisfies
    \begin{equation}
      \Re \left( \mathsf{c}_i^{(l)} \right)= \sum_{j=(i-1)b+1}^{ib}   \mathsf{e}_j^{(l)} \left(1+\zeta_j^l\right), \,\,\, |\zeta_j^l|\leq \gamma_{b-1}^l.
    \end{equation}
    At last, the sum of the computed partial sums with high-precision arithmetic satisfies
    \begin{align}
       \Re \left( \mathsf{c}^{(l)} \right) &= \sum_{i=1}^{2n/b}  \Re \left( \mathsf{c}_i^{(l)} \right)\left(1+\zeta_i^h\right), \,\,\, |\zeta_i^h|\leq \gamma_{2n/b-1}^h \\
        & = \sum_{i=1}^{2n} \mathsf{e}_i^{(l)} \left(1+\zeta_i^l\right)\left(1+\zeta_{\lceil i/b \rceil}^h\right) \\
        & = \sum_{i=1}^{2n}  \mathsf{e}_i \left(1+\delta_i\right)\left(1+\zeta_i^l\right)\left(1+\zeta_{\lceil i/b \rceil}^h\right) \\
        & = \Re \left(\mathsf{c} \right) + \Delta_1,    \label{eq:re}
    \end{align}
    where 
    \begin{equation}\small
        |\Delta_1| \leq \xi_{b,n} \sum_{i=1}^{2n}  |\mathsf{e}_i|, \, \xi_{b,n} = u_l + \gamma_{b-1}^l+\gamma_{2n/b-1}^h+O(u_l^2).
    \end{equation}

    Similarly, the imaginary part of $\mathsf{c}$ with low-precision arithmetic satisfies
    \begin{equation}
    \label{eq:im}
        \Im \left( \mathsf{c}^{(l)} \right)= \Im \left(\mathsf{c} \right) + \Delta_2,
    \end{equation}
    where 
    \begin{equation}
        |\Delta_2| \leq \xi_{b,n}\sum_{i=1}^{2n}  |\mathsf{f}_i|.
    \end{equation}
    Using \eqref{eq:re} and \eqref{eq:im}, we obtain
    \begin{equation}    
        \begin{aligned}
            \mathsf{c}^{(l)} &=  \Re \left( \mathsf{c}^{(l)} \right) +\boldsymbol{i}\Im \left( \mathsf{c}^{(l)} \right)\\
            & = \Re \left(\mathsf{c} \right) +\boldsymbol{i} \Im \left(\mathsf{c} \right) + \Delta_1 +\boldsymbol{i} \Delta_2\\
            & = \mathsf{c} + \Delta
        \end{aligned}
    \end{equation}  %\setlength\abovedisplayskip{3.5pt}\setlength\belowdisplayskip{3.5pt}
    where
    \begin{equation}\label{eq:xx}
    \begin{aligned}
            |\Delta|& = \sqrt{|\Delta_1|^2+ |\Delta_2|^2} \\
            & \leq  \xi_{b,n}  \sqrt{\left( \sum_{i=1}^{2n}  |\mathsf{e}_i|  \right)^2 + \left( \sum_{i=1}^{2n}  |\mathsf{f}_i|  \right)^2} .
    \end{aligned}
    \end{equation}
    Using Cauchy-Schwartz inequity, we have
    {\small \begin{align}
        \left( \sum_{i=1}^{2n}  |\mathsf{e}_i|  \right)^2 & \leq \sum_{i=1}^{n} \left( \Re\left(a_i \right)^2+ \Im\left(a_i \right)^2  \right) \sum_{i=1}^{n} \left( \Re\left(d_i \right)^2+ \Im\left(d_i \right)^2 \right),\\
        \left( \sum_{i=1}^{2n}  |\mathsf{f}_i|  \right)^2 & \leq \sum_{i=1}^{n} \left( \Re\left(a_i \right)^2+ \Im\left(a_i \right)^2  \right) \sum_{i=1}^{n}\left( \Im\left(d_i \right)^2 +\Re\left(d_i \right)^2 \right).
    \end{align}}
    Furthermore, \eqref{eq:xx} can be expressed as
    {\small
    \begin{align}
          |\Delta| & \leq  \xi_{b,n} \sqrt{ 2\sum_{i=1}^{n} \left( \Re\left(a_i \right)^2+ \Im\left(a_i \right)^2  \right) \sum_{i=1}^{n}\left( \Re\left(d_i \right)^2+ \Im\left(d_i \right)^2 \right)}\notag \\
            & = \sqrt{2} \xi_{b,n} \left\| \mathbf{a} \right\|_2\left\| \mathbf{d} \right\|_2.
    \end{align}}

    Therefore, \textit{Theorem} \ref{the:mixed} holds.

\bibliographystyle{IEEEtran}
\bibliography{reference}

\end{document}